\documentclass[useAMS,usenatbib]{mn2e}
\usepackage{graphicx}
\usepackage{epstopdf}

\title[Identification of Globular Cluster Stars in RAVE Data I: Application to Stellar Parameter Calibration]{Identification of Globular Cluster Stars in RAVE data I: \\ Application to Stellar Parameter Calibration}
\author[B. Anguiano et al.]{B.~Anguiano$^{1,2}$,\thanks{E-mail: borja.anguiano@mq.edu.au} D.~B.~Zucker$^{1,2,3}$, R.~-D.~Scholz$^{4}$, E.~K.~Grebel$^{5}$, G.~Seabroke$^{6}$, \newauthor A.~Kunder$^{4}$, J.~Binney$^{7}$, P.~J.~McMillan$^{7}$, T.~Zwitter$^{8}$, R.~F.~G.~Wyse$^{9}$, \newauthor G.~Kordopatis$^{4,10}$, O.~Bienaym\'e$^{11}$, J.~Bland-Hawthorn$^{12}$, C.~Boeche$^{5}$, \newauthor K.~C.~Freeman$^{13}$,  B.~K.~Gibson$^{14,15}$, G.~Gilmore$^{10}$, U.~Munari$^{16}$, J.~Navarro$^{17}$, \newauthor Q.~Parker$^{1,2,3}, $W.~Reid$^{1,2}$, A.~Siebert$^{11}$, A.~Siviero$^{18}$, M.~Steinmetz$^{4}$, F.~Watson$^{3}$ \\
\\
$^{1}$ Department of Physics and Astronomy, Macquarie University, North Ryde, NSW 2109, Australia\\
$^{2}$ Research Centre for Astronomy, Astrophysics and Astrophotonics, Macquarie University, Sydney, NSW 2109, Australia \\
$^{3}$ Australian Astronomical Observatory, North Ryde, NSW 2113, Australia \\
$^{4}$ Leibniz-Institut f\"ur Astrophysik Potsdam (AIP), An der Sternwarte 16, 14482, Potsdam, Germany\\
$^{5}$ Astronomisches Rechen-Institut, Zentrum f\"ur Astronomie der Universit\"at Heidelberg, M\"onchhofstr.\ 12--14, 69120 Heidelberg, Germany\\
$^{6}$ Mullard Space Science Laboratory, University College London, Holmbury St Mary, Dorking, RH5 6NT, UK \\
$^{7}$ Rudolf Peierls Centre for Theoretical Physics, Keble Road, Oxford OX1 3NP, UK \\
$^{8}$ Faculty of Mathematics and Physics, University of Ljubljana, Jadranska 19, 1000 Ljubljana, Slovenia \\
$^{9}$ Department of Physics and Astronomy, Johns Hopkins University, 3400 N. Charles St., Baltimore, MD 21218, USA \\
$^{10}$ Institute of Astronomy, University of Cambridge, Madingley Road, Cambridge CB3 0HA, UK\\
$^{11}$ Observatoire Astronomique de Strasbourg, 11 rue de l'Universit\'e, F-67000 Strasbourg, France \\
$^{12}$ Sydney Institute for Astronomy, University of Sydney, School of Physics A28, NSW 2006, Australia \\
$^{13}$ Research School of Astronomy and Astrophysics, Australian National University, Cotter Rd., Weston, ACT 2611, Australia\\
$^{14}$ Institute for Computational Astrophysics, Dept of Astronomy \& Physics, Saint MaryÕs University, Halifax, NS, BH3 3C3, Canada \\
$^{15}$ E.A. Milne Centre for Astrophysics, Dept of Physics \& Mathematics, Univ of Hull, HU6 7RX, UK \\  
$^{16}$ INAF Osservatorio Astronomico di Padova, 36012 Asiago (VI), Italy\\
$^{17}$ Department of Physics and Astronomy, University of Victoria, Victoria, BC, Canada\\
$^{18}$ Dipartimento di Fisica e Astronomia ÒGalileo GalileiÓ, Universita' di Padova, Vicolo dell'Osservatorio 3, I-35122 Padova, Italy}

\begin{document}

\date{Accepted, 3 April 2015}

\pagerange{\pageref{firstpage}--\pageref{lastpage}} \pubyear{2015}

\maketitle

\label{firstpage}

\begin{abstract}

We present the identification of potential members of nearby Galactic globular clusters using radial velocities from the RAdial Velocity Experiment Data Release 4 (RAVE-DR4) survey database. Our identifications are based on three globular clusters -- NGC 3201, NGC 5139 ($\omega$ Cen) and NGC 362 -- all of which are shown to have $\mid$RV$\mid$ $>$ 100 km s$^{-1}$. The high radial velocity of cluster members compared to the bulk of surrounding disc stars enables us to identify members using their measured radial velocities, supplemented by proper motion information and location relative to the tidal radius of each cluster. The identification of globular cluster stars in RAVE DR4 data offers a unique opportunity to test the precision and accuracy of the stellar parameters determined with the currently available Stellar Parameter Pipelines (SPPs) used in the survey, as globular clusters are ideal testbeds for the validation of stellar atmospheric parameters, abundances, distances and ages. 
For both NGC 3201 and $\omega$ Cen, there is compelling evidence for numerous members ($> 10$) in the RAVE database; in the case of NGC 362 the evidence is more ambiguous, and there may be significant foreground and/or background contamination in our kinematically-selected sample. A comparison of the RAVE-derived stellar parameters and abundances with published values for each cluster and with BASTI isochrones for ages and metallicities from the literature reveals overall good agreement, with the exception of the apparent underestimation of surface gravities for giants, in particular for the most metal-poor stars. Moreover, if the selected members are part of the main body of each cluster our results would also suggest that the distances from \citet{2013MNRAS.tmp.2584B}, where only isochrones more metal-rich than -0.9 dex were used, are typically underestimated by $\sim$ 40$\%$ with respect to the published distances for the clusters, while the distances from \citet{2010A&A...522A..54Z} show stars ranging from 1 to $\sim$ 6.5 kpc -- with indications of a trend toward higher distances at lower metallicities -- for the three clusters analysed in this study.   


\end{abstract}


\section{Introduction}

In the era of massive stellar spectroscopic surveys, automated Stellar Parameter Pipelines (SPPs) and their validation are crucial for the scientific exploitation both of existing Galactic surveys, such as SEGUE \citep{2009AJ....137.4377Y} and RAVE \citep{2006AJ....132.1645S}, and those in progress, such as \emph{Gaia}, which will measure spectra for $\sim$ 150 million stars \citep{2012Ap&SS.341...31D}, Gaia-ESO \citep{2012Msngr.147...25G}, APOGEE \citep{2008AN....329.1018A}, and GALAH \citep{2008ASPC..399..439F, 2013AAS...22123406Z, 2014IAUS..298..322A,2015MNRAS...in press}, where observations of a million stars are planned. However, there are limited opportunities for checking the outputs of these automated SSPs against more traditional analyses in the literature, aside from dedicated observations of reference or calibration stars.

Galactic globular cluster (hereafter GGC) members offer a unique opportunity to validate the precision and accuracy of fundamental stellar atmospheric parameters obtained using currently available SPPs \citep{2008AJ....136.2050L, 2011AJ....141...89S}. The GGC population in the halo of the Milky Way covers a wide range of metal abundances, essentially independent of radius from the Galactic Centre, spanning approximately -0.5 dex to -2.2 dex. In addition these objects have an age spread of 2 - 3 Gyrs \citep{2009ApJ...694.1498M, 2013ApJ...775..134V}, being mostly older than 10 Gyr \citep{1996AJ....112.1487H}. However, the traditional paradigm treating GGCs as single stellar populations has largely fallen by the wayside in recent years. Multiple generations of stars have been detected from photometry and spectroscopy in a number of GGCs \citep{2007ApJ...661L..53P, 2010ApJ...709.1183M, 2012A&ARv..20...50G}. For a number of massive star clusters, like $\omega$ Cen, several distinct episodes of star formation have been discovered \citep{2010ApJ...722.1373J}. Very massive star clusters are considered as possible cores or nuclei of stripped dwarf galaxies \citep{2003MNRAS.346L..11B}. High-resolution spectroscopic studies of individual stars in GGCs have revealed that some of these objects have a substantial star-to-star metallicity scatter. \citet{2011A&A...532A...8M} reported a range in metallicity from -2.0 dex to -1.6 dex in a data-set of 35 red giants in M22. \citet{2010A&A...520A..95C}, using high-resolution spectra of 76 red giants, found that the bulk of stars peak at [Fe/H] $\sim$ -1.6, with a long tail extending to higher metallicities, in the globular cluster M54. Very precise abundance determinations for several members of NGC 3201 with high-resolution spectroscopy show a possible metallicity spread of 0.12 dex is present in the cluster \citep{2013MNRAS.433.2006M}. However, except for variations in their light element abundances \citep{2010ApJ...712L..21C}, most GGCs seem to be mono-metallicity objects; that is, they have roughly uniform iron abundances. A real spread in metallicity seems to be rare.

The period of time over which chemically distinct multiple generations of stars in GGCs are believed to have formed -- $\sim 10^{8}$ to $\sim 10^{9}$ years -- \citep{2011ApJ...726...36C, 2012A&ARv..20...50G} is still one to two orders of magnitude shorter than the history of star formation in the Galactic disc, and these stars formed within a relatively small volume (r $\sim$ tens of pc). Hence any identified GGC members in the RAVE survey database would serve as excellent test subjects for validating the estimated distances and ages in the RAVE survey using stellar atmospheric parameters; the typical uncertainty in the distances to globular clusters is $\sim$ 6$\%$, which leads to a 13$\%$ uncertainty in the absolute ages \citep{2009IAUS..258..221S, 2013ApJ...775..134V}. In this paper, we report on the identification of members of nearby globular clusters in the RAVE catalogue, and use these identifications, in conjunction with the properties of these clusters published in the literature, to test the basic stellar properties obtained for these stars from the RAVE survey. (An independent search of RAVE data for GGC members, with the goal of using new extra-tidal stars as tracers of the clusters' disruption and possible accretion origins, has been carried out by \citealt{2014arXiv1408.6236K}.)

This paper is organised as follows. In Section 2 we describe the RAVE survey. Cluster membership selection is described in Section 3. In Section 4 we use the likely cluster members to test the stellar parameters in the RAVE survey. We present our conclusions in Section 5.

\begin{figure*}
  \centering  
  \includegraphics[width=1.99\columnwidth]{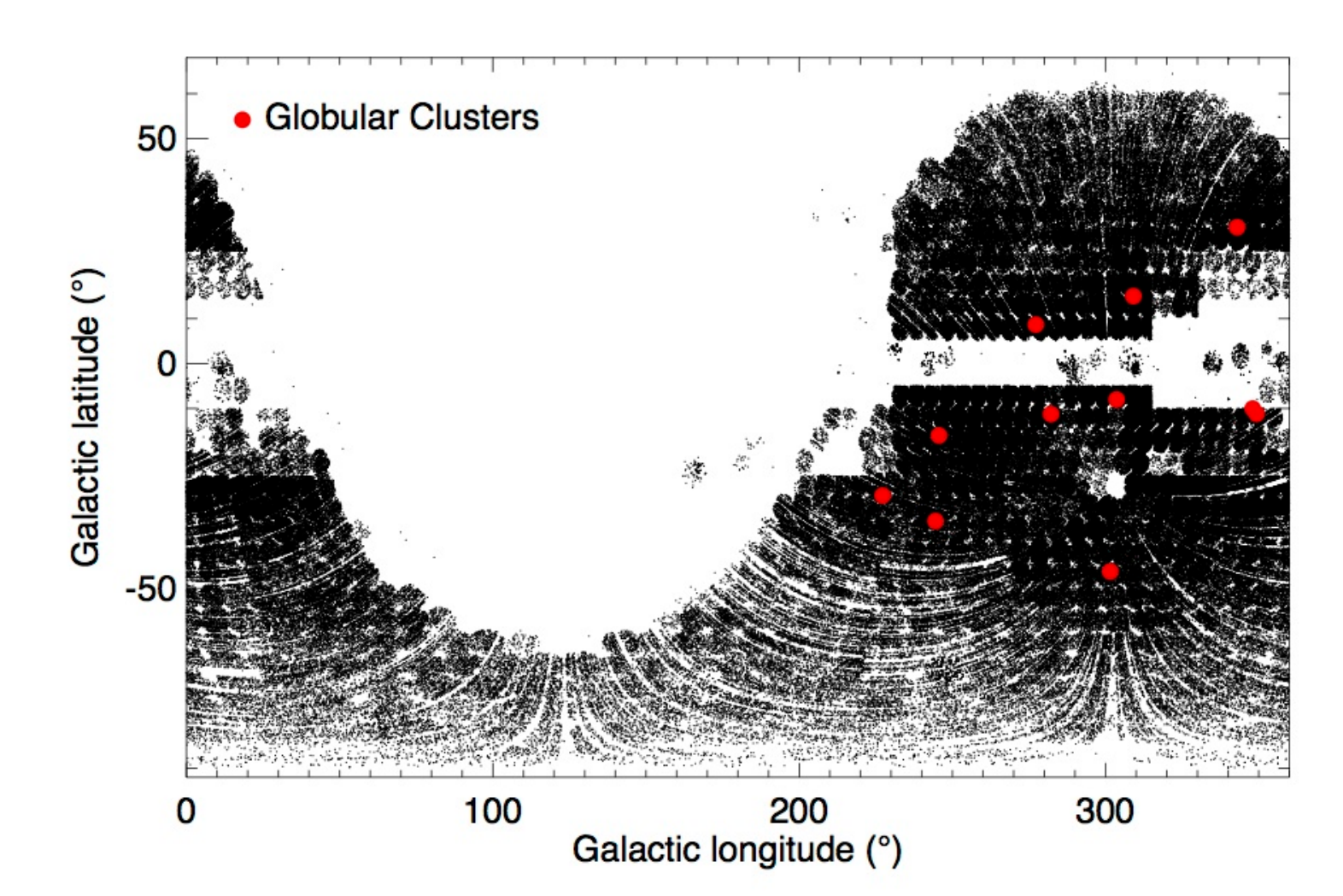}
  \caption{Galactic coordinates plot of all the stars observed in the RAVE survey. The large red dots indicate the positions of globular clusters for which we identified potential members in the survey database.}
  \label{fig:example}
\end{figure*}

\section{The RAVE survey}

The RAdial Velocity Experiment (RAVE; \citealt{2006AJ....132.1645S}) is a spectroscopic survey which used the Six Degree Field (6dF) multi-object spectrograph on the 1.2m UK Schmidt Telescope of the Australian Astronomical Observatory (AAO) at Siding Spring Observatory (SSO). Stars were initially drawn from the pilot survey input catalogue based on Tycho-2 \citep{2000A&A...355L..27H} and SuperCOSMOS \citep{2001MNRAS.326.1279H} before the main input catalogue based on DENIS \citep{1997Msngr..87...27E}, in the magnitude range 9 $<$ I $<$ 13. The RAVE survey provides, via the fourth data release (DR4), the radial velocities and stellar atmospheric parameters for 483,849 objects, derived using medium-resolution spectra (R = 7,500) in the Ca II triplet region (8410 - 8795 \AA). RAVE data are complemented with proper motions from Tycho-2, UCAC2, UCAC3, PPMX, PPMXL and SPM4, in addition to photometric data from the major optical and infrared catalogs, Tycho-2, USNO-B, DENIS, 2MASS and APASS \citep{2006AJ....132.1645S, 2008AJ....136..421Z, 2011AJ....141..187S, 2013AJ....146..134K}. \citet{2011AJ....142..193B} presented the RAVE chemical catalogue. It contains chemical abundances for the elements Mg, Al, Si, Ca, Ti, Fe, and Ni, with a mean error of $\sim$0.2 dex, as judged from accuracy tests performed on both synthetic and real spectra. \citet{2010A&A...511A..90B} developed a method for estimating distances from RAVE spectroscopic data, stellar models and (J-Ks) photometry from archival sources to derive absolute magnitudes. \citet{2010A&A...522A..54Z} determined new distances with a method assuming that the star undergoes a standard stellar evolution and that its spectrum shows no peculiarities. \citet{2010MNRAS.407..339B} applied the Bayesian scheme of \citet{2010MNRAS.407..339B2} to the DR3 data. Recently, \citet{2013MNRAS.tmp.2584B} utilised the \citet{2010MNRAS.407..339B2} technique on the DR4 parameters with the addition of H-band photometry and a determination of the extinction to estimate stellar distances. 
  
\section{The sample}

We identified potential stellar members for three GGCs using the RAVE catalogue. The clusters are NGC 3201, NGC 5139 ($\omega$ Cen) and NGC 362. We note that we also have identified potential stars from numerous other clusters -- including NGC 2298, NGC 2808, NGC 4833, NGC 5897, NGC 6496, NGC 6541, NGC 1904 (M79) and NGC 1851 --  in the RAVE database, but, as the systemic velocities of these clusters are not sufficiently separated from the velocity distribution of Galactic foreground stars, we cannot determine likely membership without taking abundances into account, thereby introducing a metallicity bias into our selection criteria. Hence we restrict our analysis in this study to the aforementioned three kinematically distinct GGCs. 

Table 1 summarises the main properties of each of the clusters included in this study. Positions and radial velocities come from the compilation of the \citet{1996AJ....112.1487H} catalogue (2010 edition) while the absolute proper motions of globular clusters come from \citet{2007AJ....134..195C, 2010AJ....140.1282C, 2013AJ....146...33C} and references therein. Fig.~1 shows the positions of all the stars observed in the RAVE survey in Galactic coordinates. The red points indicate the positions of the GGCs found in this study.

\begin{table*}
 \centering
 \begin{minipage}{165mm}
  \caption{Properties from the \citet{1996AJ....112.1487H} catalogue (2010 edition) of GGCs identified in the RAVE data with $\mid$RV$\mid$ $>$ 100 km s$^{-1}$. Distance uncertainties are consistent with a $\pm$ 0.15 dex change in the distance modulus.}
  \begin{tabular}{@{}llrrrrlrlr@{}}
  \hline
   ID & RA(hh:mm:ss) & DEC(dd:mm:ss) & l ($^\circ$) & b ($^\circ$) & V$_{r}$ (km/s) &  $\mu_{\alpha}$ (mas/yr) & $\mu_{\delta}$ (mas/yr) & Dist. \\
    & (J2000)  & (J2000) &  &  & &  & & (kpc)  \\
 \hline
 NGC 362   & 01:03:14.26  & -70:50:55.6  & 301.53  & -46.25  & +223.5  $\pm$ 0.5 & +4.873 $\pm$ 0.514  & -2.727  $\pm$  0.824  & 8.6 $\pm$ 0.6 \\
 NGC 3201 & 10:17:36.82  & -46:24:44.9  & 277.23  & +8.64   & +494.0 $\pm$ 0.2  & +5.280 $\pm$ 0.320  & -0.980  $\pm$  0.330  & 4.9 $\pm$ 0.3  \\
 NGC 5139 & 13:26:47.24  & -47:28:46.5  & 309.10  & +14.97 & +232.1  $\pm$ 0.1 & -5.080  $\pm$ 0.350  & -3.570  $\pm$  0.340  & 5.2 $\pm$ 0.3\\ 
 \hline
\end{tabular}
\end{minipage}
\end{table*}


\subsection{Cluster membership selection}

We selected probable cluster members based on their measured radial velocities (RVs) and absolute proper motions. RAVE RVs are computed by cross-correlation of sky-subtracted normalised spectra with an extensive library of synthetic spectra. Spectra without sky subtraction are used to compute the zero-point correction. The typical RV accuracy for RAVE data is $\leq$ 2 km s$^{-1}$. For measurements with a high signal-to-noise ratio (S/N), the error is only 1.3 km s$^{-1}$, with a negligible zero-point error \citep{2006AJ....132.1645S, 2008AJ....136..421Z, 2011AJ....141..187S, 2013AJ....146..134K}. 

\begin{figure}
  \centering  
  \includegraphics[width=0.99\columnwidth]{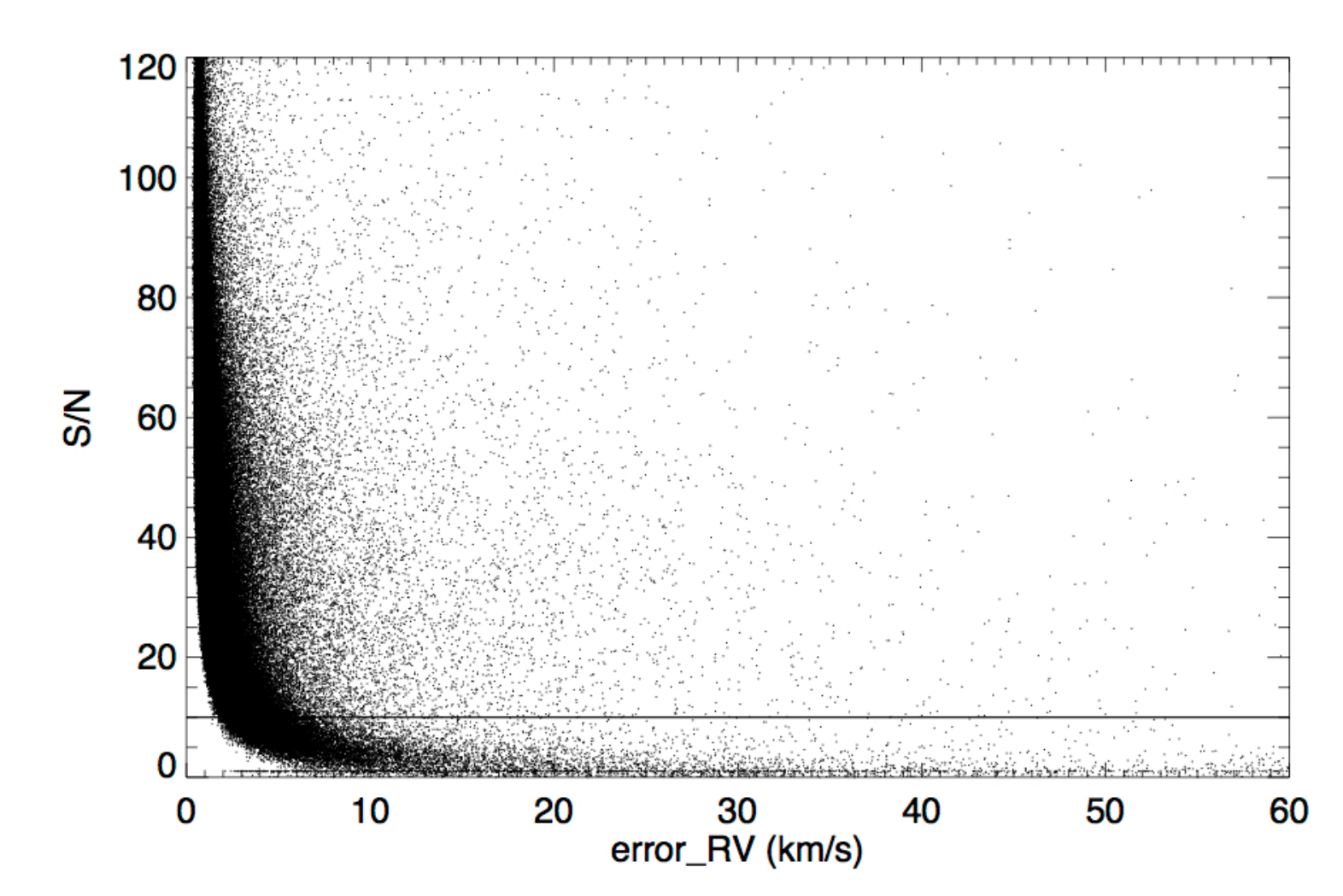}
  \caption{RAVE radial velocity measurement errors with respect to the S/N of the spectrum. For this study we selected stars with S/N $>$ 10 (indicated by a solid line in the plot), where $\sim$ 80$\%$ have a $\sigma_{RV} \leq$ 5 km s$^{-1}$.}
  \label{fig:erRV_SN}
\end{figure}

In Fig.~\ref{fig:erRV_SN} we present the behaviour of the error in the RV with the S/N of the spectra. (For more detail on S/N measurements in the RAVE spectra see \citealt{2008AJ....136..421Z}.)  In this work we selected stars with S/N $>$ 10 (black line in figure); these stars have a typical internal RV error $\leq$ 5 km s$^{-1}$. Note that RAVE spectra with low S/N (i.e. S/N $<$ 10) and stars with spectral peculiarities \citep{2012ApJS..200...14M} typically have lower precision RVs. Fig.~\ref{fig:erRV_SN} shows that the typical accuracy of the RV increases with the S/N, with error in RV $<$ 2 km s$^{-1}$ for 90$\%$ of the catalogue at S/N $\sim$ 100. We also selected stars with a \citet{1979AJ.....84.1511T} cross-correlation coefficient R larger than 5. RAVE provides very precise RVs for 80$\%$ of the catalogue. For this reason our main criterion for membership selection was a comparison of the stellar RVs in the area of the sky around each GGC to the systemic RV for that GGC from the literature (see Table~1). 

In this paper we restricted our study to the three clusters that have $\mid$RV$\mid$ $>$ 100 km s$^{-1}$, in order to get cleaner samples of halo stars relatively uncontaminated by disc stars. This purely kinematically derived sample does not require us to make any cuts in metallicity for membership selection, thereby leaving abundances as a free parameter for directly testing the reliability of the RAVE SPP. Once candidates were selected as explained above we used the tidal radius of the clusters and the proper motions from the PPMXL catalogue \citep{2010AJ....139.2440R} to select the highest likelihood members for the final dataset.         

\begin{table}
 \centering
 \begin{minipage}{75mm}
  \caption{Structural parameters of Galactic Globular Clusters identified in the RAVE data from the \citet{1996AJ....112.1487H} catalogue (2010 edition).}
  \begin{tabular}{@{}llrrrrlrlr@{}}
  \hline
   ID & Central $\sigma{_V}$ & r$_{c}$ & r$_{t}$ \\
   & (km/s) & (arcmin) & (arcmin) \\
 \hline
 NGC3201 &   5.0 $\pm$ 0.2  & 1.30  & 25.34  \\
 NGC5139 & 16.8 $\pm$ 0.3  & 2.37  & 48.38  \\
 NGC362   &   6.4 $\pm$ 0.3  & 0.18  & 10.35  \\
\hline
\end{tabular}
\end{minipage}
\end{table}


\subsubsection{Selection of potential members based on radial velocities}

\begin{figure*}
  \centering  
  \includegraphics[width=0.99\columnwidth]{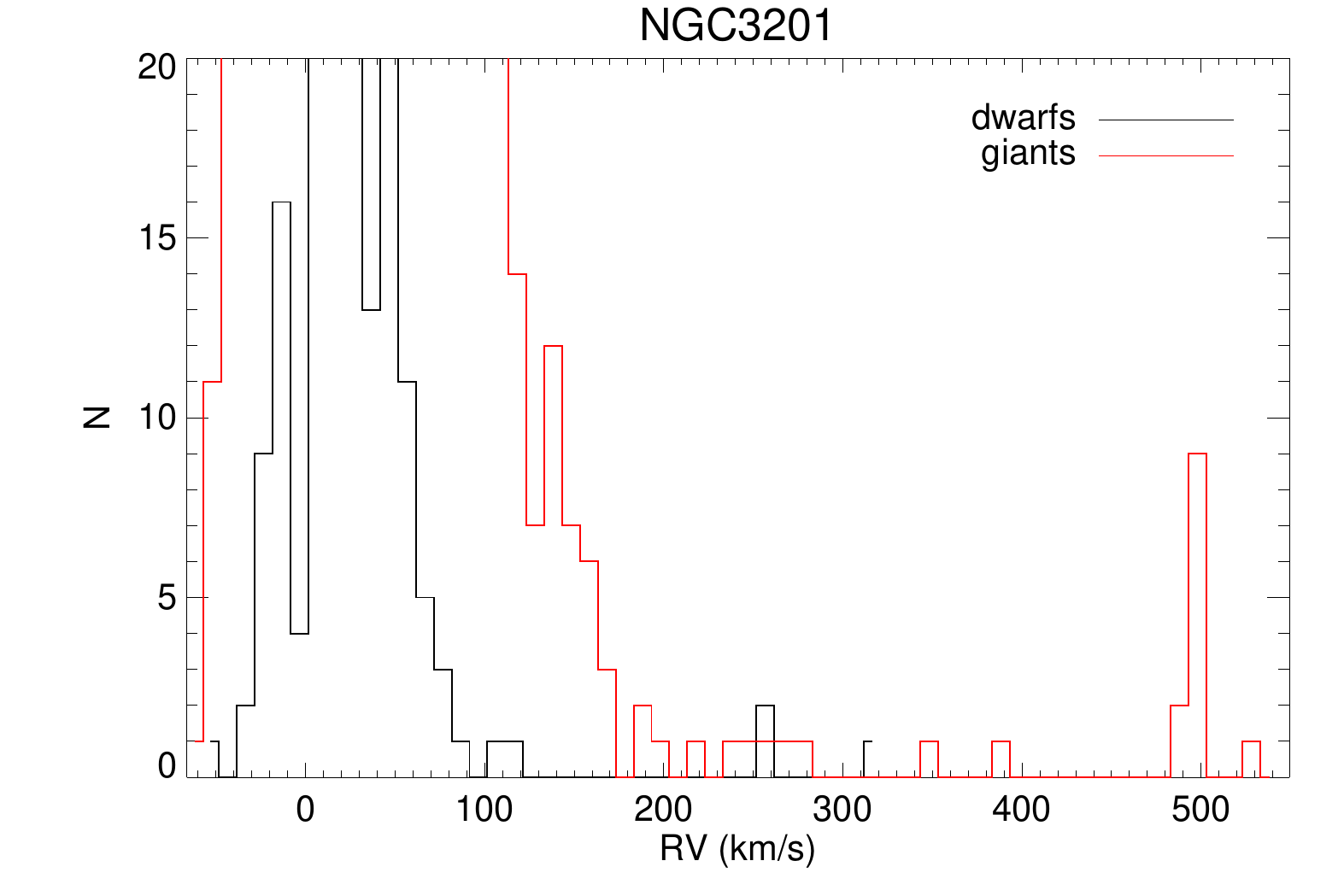}
  \includegraphics[width=0.99\columnwidth]{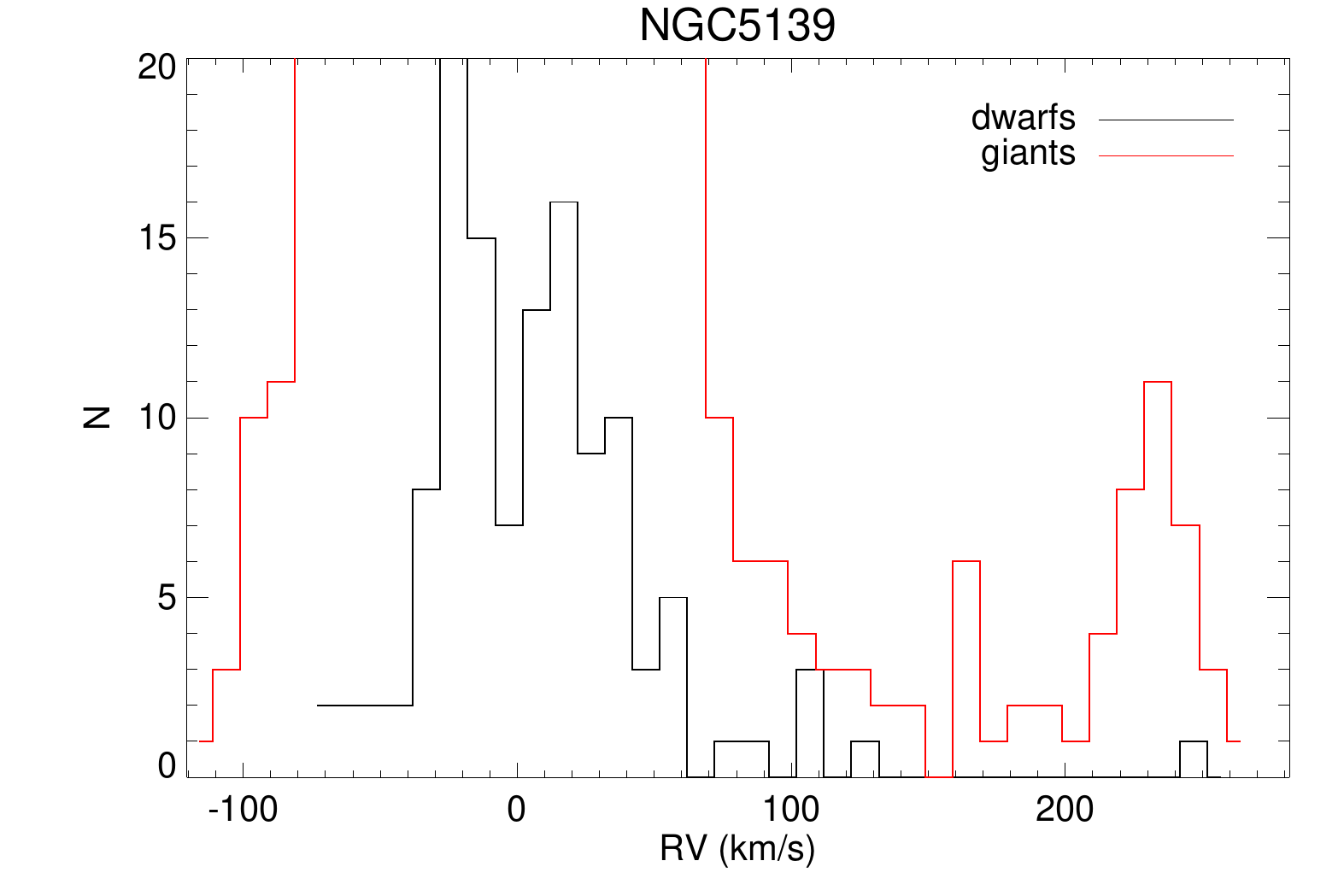}
  \includegraphics[width=0.99\columnwidth]{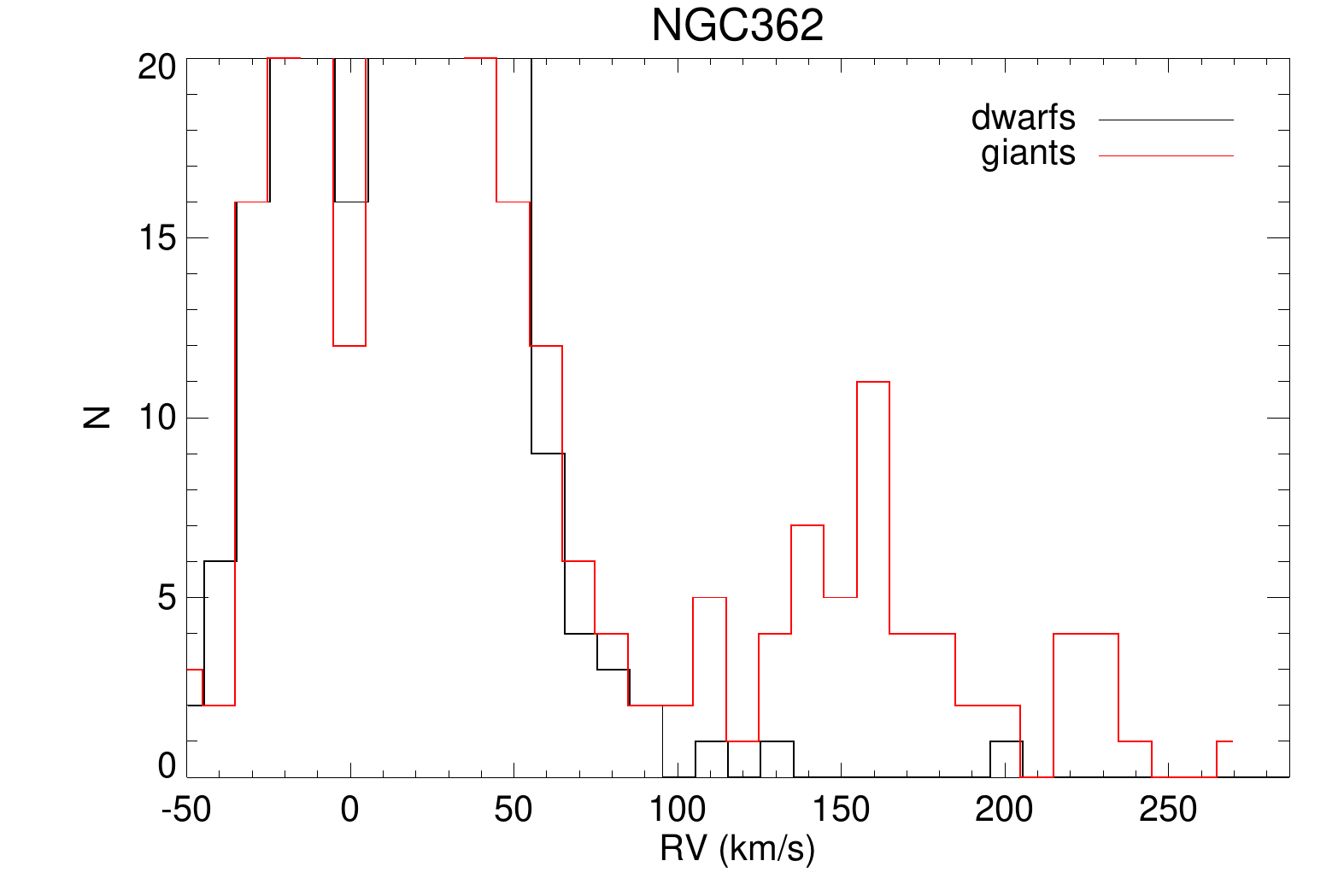}
  \caption{Radial Velocity Distribution Functions (RVDFs) for an area of 6 square degrees centred at the positions of NGC 3201, NGC 5139 ($\omega$ Cen) and NGC 362. The black line represents dwarf stars and the red line giants (see text for an explanation of the selection of dwarfs and giants). Note the peak around 490 km s$^{-1}$ related with NGC 3201, around 230 km s$^{-1}$ associated with the RV for NGC 5139 and around 230 km s$^{-1}$ for NGC 362. The distribution of stars between 120 $<$ RV $<$ 200 km s$^{-1}$ in the panel for NGC 362 is intriguing; they may be associated with the Small Magellanic Cloud (SMC).}
  \label{fig:RVDF}
\end{figure*}

Nearby GGCs that have a large RV ($\mid$RV$\mid$ $>$ 100 km s$^{-1}$) can serve as unbiased abundance calibrators for stellar surveys, as these objects can be identified in the RV Distribution Function (RVDF hereafter) from the bulk of the Milky Way's stellar disc for the area in the sky surrounding the cluster. RAVE is a magnitude-limited survey covering 9 to 13 in the I band. The three clusters we consider here are a few kpc away; in this magnitude range, the candidate members of one of these clusters will likely belong to the upper parts of the red and asymptotic giant branches, i.e., the brightest giants of a cluster. Fig.~\ref{fig:RVDF} shows the RVDF for a 6 square-degree area centred on each cluster, in which we have identified potential members in the RAVE data. We made a broad selection between dwarfs and giants using RAVE-derived surface gravities and temperatures (which, owing to their different luminosity classes, probe different volumes). The black line represents stars in the range 3500 K $<$ T$_{\rm eff}$ $<$ 8000 K and log g $>$ 3.5 (cgs) while the red line represents stars in the range 3600 K $<$ T$_{\rm eff}$ $<$ 6000 K and log g $<$ 3.5 (cgs). 

{\bf NGC 3201} has an extremely high radial velocity (RV $\sim$ +494 km s$^{-1}$), which makes this cluster an ideal target to test our selection method. High-velocity halo star studies using the RAVE survey have shown that the number of stars with a RV larger than $\pm$400 km s$^{-1}$ is not very large \citep{2007MNRAS.379..755S, 2014A&A...562A..91P}. In Fig.~3 we identify a group of eleven stars with a RV $\sim$ 495 km s$^{-1}$ in the area of the sky around NGC 3201. This group of stars is clearly distinct in velocity from the bulk of the stars in this region of sky, and, given their similar RV values, they are very likely members of NGC 3201. The central velocity dispersion for this cluster is only 5 km s$^{-1}$ (see Table~2), which is greater than the mean RV error ($<$ 3 km s$^{-1}$).

Candidates for the globular cluster {\bf NGC 5139} ($\omega$ Cen) also appear clearly in the RAVE data. In the RVDF there is a peak of stars around 230 km s$^{-1}$, exactly the RV reported for this cluster (see Table~1). As mentioned above, $\omega$ Cen is a massive, complex system with multiple star formation episodes \citep{2013ApJ...762...36J}. \citet{2008AJ....136..506D}, \citet{2010AJ....139..636W} and \citet{2012ApJ...747L..37M} detected field stars that may be associated with $\omega$ Cen in the nearby Galactic disc. We found a broad RV distribution around the nominal RV for this cluster, from $\sim$ 200 to 260 km s$^{-1}$ (Fig.~3); however, it is important to note that this cluster has a significant central internal dispersion, $\sigma_{V}$ = 16.8 km s$^{-1}$. Recently, \citet{2012ApJ...751....6D} measured a line-of-sight velocity dispersion in the outer parts of $\omega$ Cen of $\sim$ 6.5 km s$^{-1}$.

The case for identifying members of  {\bf NGC 362} is less clear-cut in the RAVE data. This cluster has an RV $\sim$ +223 km s$^{-1}$. The RVDF in Fig.~3 shows only three potential candidates for NGC362 membership. We also find a potential stellar radial velocity over-density between 100 $<$ RV $<$ 200 km s$^{-1}$; some of these stars may be associated with the Small Magellanic Cloud (SMC) \citep{1998AJ....115.1934D}.

\subsubsection{Selection of probable members using tidal radii and proper motions}

As a consistency check to verify the validity of our probable cluster members selected using RVs only, here we check the proper motions of the candidates as well as their position within the clusters' tidal radii.

The central potential of the cluster defines the density distribution within the cluster, where the radius is in units of the King radius, r$_{0}$ \citep{1966AJ.....71...64K}. At the radius where the potential falls to zero, the density also falls to zero. This is the so-called \emph{tidal radius}, r$_{\rm t}$, and it is the radius at which the inward force towards the cluster centre is balanced by the outward pull of the Galaxy's tidal field, so that at $r>r_{\rm t}$ stars are not bound to the cluster. The central concentration parameter of a King model is defined to be

\begin{equation}
c \equiv \log_{10} (r_{\rm t}/r_0) 
\end{equation}

where $r_0$ is the radius at which the projected surface brightness falls to half its central value. Table 2 lists the core radius and the tidal radius for the globular clusters presented in this study, as well as the internal velocity dispersion of the clusters. 

\begin{figure}
  \centering  
  \includegraphics[width=1.02\columnwidth]{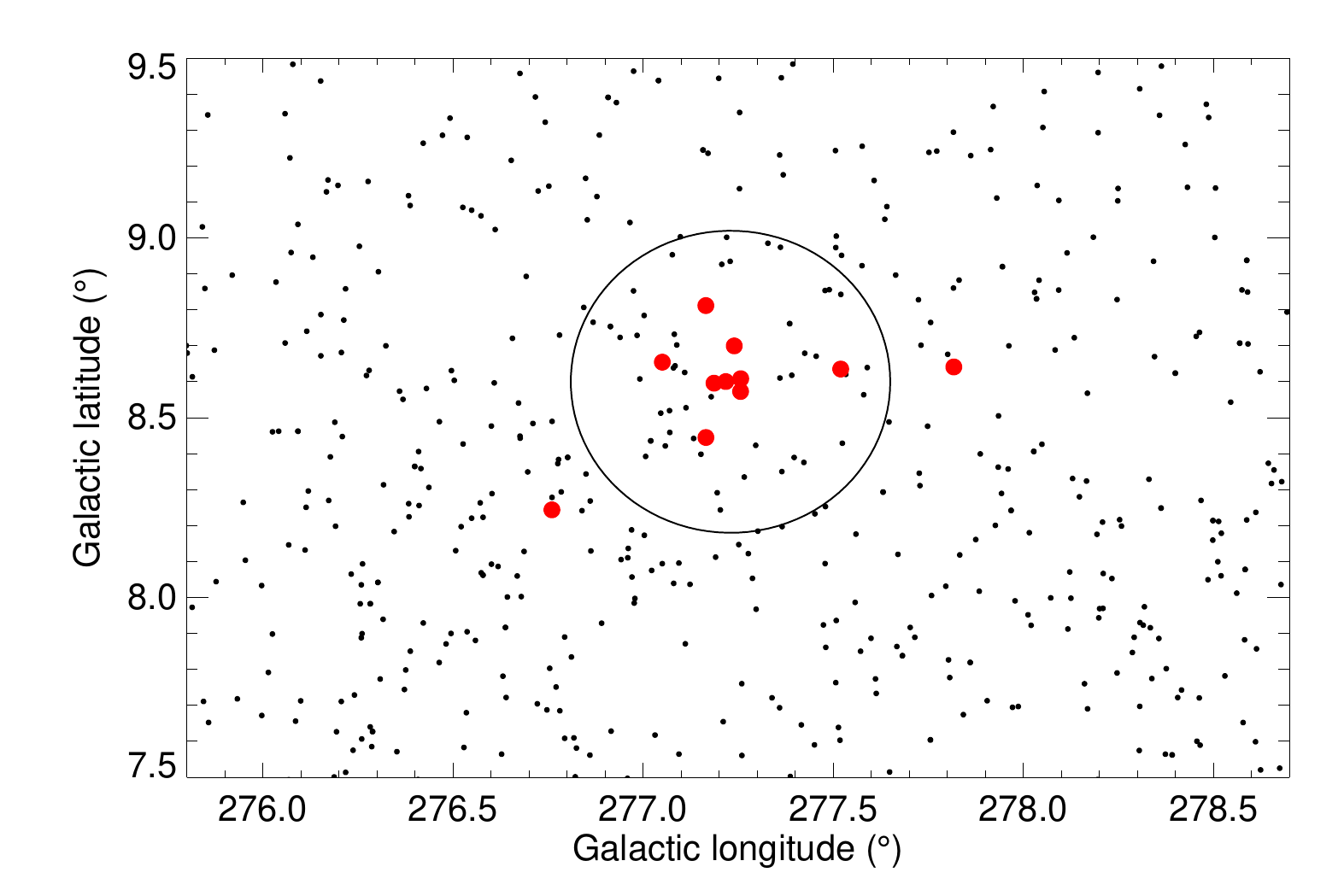}
  \includegraphics[width=1.02\columnwidth]{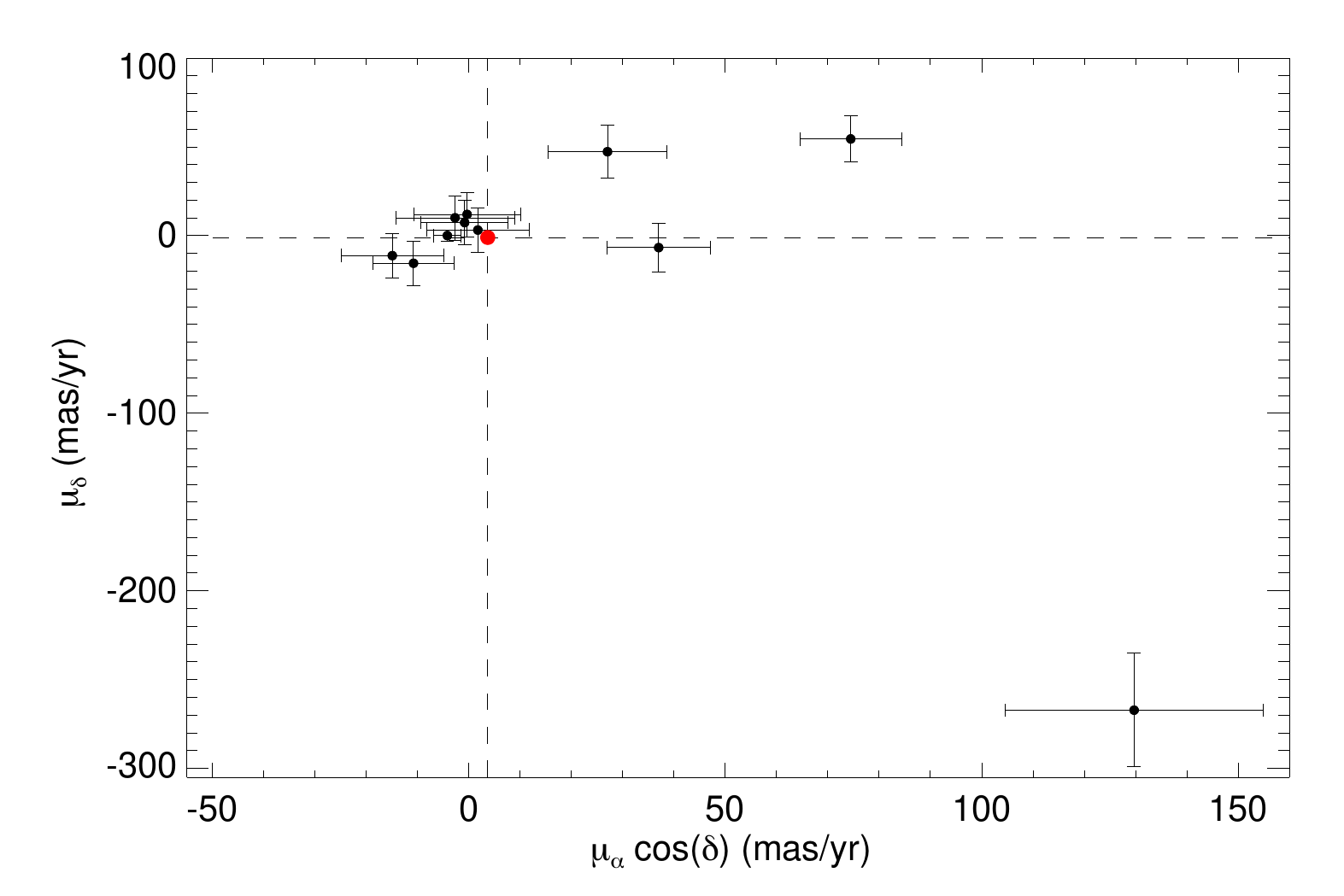}
  \caption{\emph{Top panel:} The positions (in Galactic coordinates) for the stars in the RAVE catalogue in the region of NGC 3201. The red dots are potential members of NGC 3201 selected based on RV. The black circle is the tidal radius according to the value given in \citet{1996AJ....112.1487H} (2010 edition) for this cluster. \emph{Bottom panel:} The proper motion plane (mas/yr) for the selected candidates. The red dot indicates the absolute proper motion of the cluster \citep{2007AJ....134..195C, 2010AJ....140.1282C, 2013AJ....146...33C}.}
  \label{fig:NGC3201_pos}
\end{figure}
 
For the cluster {NGC 3201} we selected eleven stars using RV alone. Nine of these stars are inside the tidal radius (see Fig. \ref{fig:NGC3201_pos}) with proper motions similar to the nominal value for the cluster ($\mu_{\alpha}$ cos($\delta$) = 3.65,  $\mu_{\delta}$ = 0.98). We also find four stars with significantly higher proper motion values, suggesting that these stars may be foreground objects. 

\begin{table*}
 \centering
 \begin{minipage}{370mm}
  \caption{NGC 3201 candidates selected from RAVE data and their parameters. The last column indicates the \emph{crowding} (see Section 3.2).}
  \begin{tabular}{@{}rrrrrrrrrrr@{}}
  \hline
  \hline
ID & R.A. & Decl. & RV & $\mu_{\alpha}$ cos($\delta$) & $\sigma_{\mu_{\alpha}}$ & $\mu_{\delta}$ & $\sigma_{\mu_{\delta}}$ & \emph{J} & S/N & C.  \\
   &&& (km/s) & (mas/yr) & (mas/yr) & (mas/yr) & (mas/yr) & \\
 \hline
J101405.6-462841  &10h 14' 05.63'' & -46$^{\circ}$ 28' 40.6'' &   494.56 $\pm$   0.62 &     1.76    &  12.00&   3.20       & 12.00 & 10.01& 35.9 & -\\
J101640.5-463221 &10h 16' 40.52''  & -46$^{\circ}$ 32' 20.5'' &   493.94 $\pm$   0.83 &   -0.33     &  12.00&  12.00      & 12.00 & 10.14& 34.5 & u\\
J101648.9-461807 &10h 16' 48.88''  & -46$^{\circ}$ 18' 06.9'' &   496.44 $\pm$   0.91 &   -0.82    &  12.00&   7.40     &  12.00 &  9.16 &  33.7  & -\\
J101716.2-462533 &10h 17' 16.16'' & -46$^{\circ}$ 25' 32.8'' &   502.19 $\pm$   2.57 &    74.46   &  12.00&   54.50     & 12.00 &  9.26  & 45.7 & v, u\\
J101725.9-462621 &10h 17' 25.86''  & -46$^{\circ}$ 26' 21.0'' &   489.22 $\pm$  1.05 &    27.03  &  14.10 & 47.30    &  14.00 & 8.69  &  23.1 & v, u\\
J101731.6-462901 &10h 17' 31.59''  & -46$^{\circ}$ 29' 01.0'' &   500.99 $\pm$   1.67 &   -14.91  &  12.00&  -11.20   &  12.00 & 10.67&  22.0  & -\\
J101738.6-462716 &10h 17' 38.59'' & -46$^{\circ}$ 27' 16.0'' &   494.65 $\pm$   1.41 &   129.70 &  32.00&  -267.20 &  32.00 &     -    &  36.5  & v, u\\
J101751.5-462210 &10h 17' 51.53'' & -46$^{\circ}$ 22' 09.7'' &   495.87 $\pm$   0.68 &    37.01  &  13.00&   -6.70     & 13.00 &  9.47 &  46.8 & v, u\\ 
J101752.1-461407  &10h 17' 52.04'' & -46$^{\circ}$ 14' 06.6'' &   483.86 $\pm$   0.61 &   -10.78  &  12.00&  -15.50    & 12.00 &  9.71 &  50.6  & -\\ 
J101859.1-463438 &10h 18' 59.10'' & -46$^{\circ}$ 34' 37.6'' &   497.94 $\pm$   0.86 &   -4.18    &   3.00 &   0.10     &    3.01 & 8.44  &  43.0  & -\\
J102025.9-464406 &10h 20' 25.86'' & -46$^{\circ}$ 44' 05.8'' &   496.77 $\pm$  0.69 &   -2.68    &  12.00 & 10.10    &  12.00 & 8.41  &  40.8  & - \\
\hline
\label{tab:NGC3201_cands}
\end{tabular}
\end{minipage}
\end{table*}

Table \ref{tab:NGC3201_cands} summarises the candidates of NGC3201. The two stars outside of the tidal radius are J102025.9-46440610 and J101405.6-46284110; however both of these stars have similar proper motions to the cluster, and hence are still candidates for membership. The stars with high proper motions (J101725.9-462621, J101738.6-462716, J101751.5-46221010 and J101716.2-462533) could be catalogued as field stars despite being inside the tidal radius. Despite the marked differences in proper motion compared to NGC 3201, the striking similarity in RV makes these stars interesting targets. A detailed chemical abundances analysis is necessary to understand their relation, if any, with the cluster. All the stars have spectra with a S/N $>$ 30, except for two with S/N $>$ 20.

\begin{figure}
  \centering  
  \includegraphics[width=1.02\columnwidth]{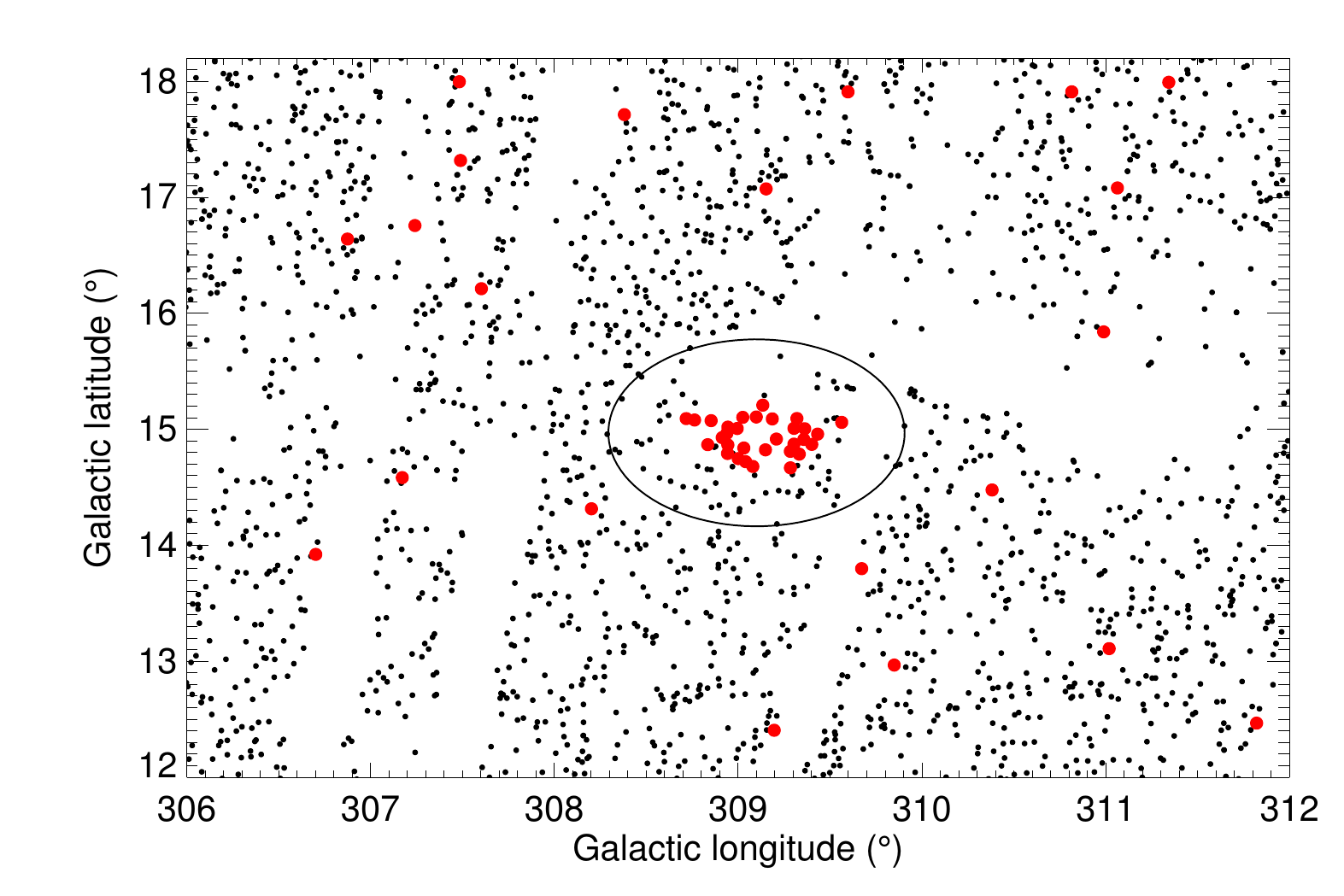}
  \includegraphics[width=1.02\columnwidth]{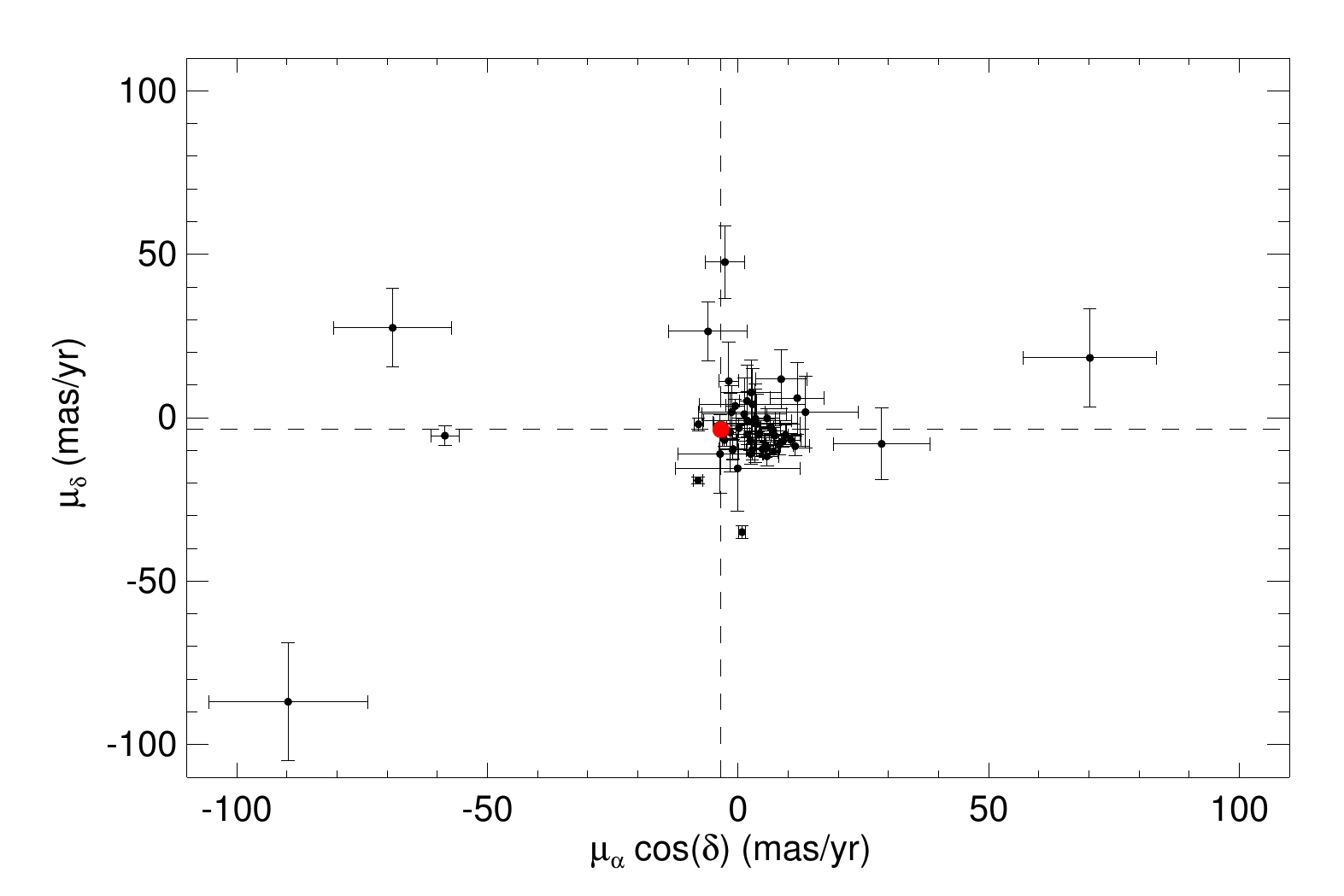}
  \caption{\emph{Top panel:} The positions (in Galactic coordinates) for the stars in the RAVE catalogue in the region of $\omega$ Cen. The red dots are potential members of $\omega$ Cen selected based on RV. The black circle is the tidal radius according to the value given in \citet{1996AJ....112.1487H} (2010 edition) for this cluster. We identify several stars outside of the tidal radius as candidate cluster members. \emph{Bottom panel:} The proper motion plane (mas/yr) for the selected candidates. The red dot indicates the absolute proper motion of the cluster \citep{2007AJ....134..195C, 2010AJ....140.1282C, 2013AJ....146...33C}.}
  \label{fig:NGC5139_pos}
\end{figure}

\begin{table*}
 \centering
 \begin{minipage}{370mm}
  \caption{NGC 5139 ($\omega$ Cen) candidates selected from RAVE data and their parameters. The last column indicates the \emph{crowding}.}
  \begin{tabular}{@{}rrrrrrrrrrrrr@{}}
  \hline
  \hline
ID & R.A. & Decl. & RV & $\mu_{\alpha}$ cos($\delta$) & $\sigma_{\mu_{\alpha}}$ & $\mu_{\delta}$ & $\sigma_{\mu_{\delta}}$ & \emph{J} & S/N & C.  \\
   &&& (km/s) & (mas/yr) & (mas/yr) & (mas/yr) & (mas/yr)  \\
 \hline
J131313.9-460352 & 13h 13' 13.88" &   -46$^{\circ}$ 03' 52.0"   &   180.59  $\pm$  0.84   & 11.79  &  11.00  &  6.00   &  11.00   &  10.33 & 45.1 & -\\
J131340.4-484714 & 13h 13' 40.43" &   -48$^{\circ}$ 47' 13.8"   &   220.25  $\pm$   1.07  & -1.58  & 12.00   & -4.50   &  12.00    & 10.63 & 33.0 & -\\
J131511.1-455458 & 13h 15' 11.06"  &  -45$^{\circ}$ 54' 57.5"  &   165.54   $\pm$   1.88  & -2.63   &   11.00  &   47.60  &   11.00  & 10.65  & 14.8 & v, u \\
J131548.1-443935 & 13h 15' 48.07" &   -44$^{\circ}$ 39' 34.7"  &    314.76  $\pm$   0.59   & 7.09 &  1.00 &  -10.40 &  1.00 &  9.50 & 57.3 & -\\
J131602.3-480507 & 13h 16' 02.28" &   -48$^{\circ}$ 05' 06.6"   &   257.20  $\pm$  0.95   & 8.59   &  9.00    & 11.80   &  9.00 &  9.66 & 21.1 & -\\
J131613.1-452004 & 13h 16' 13.13" &   -45$^{\circ}$ 20' 03.7"   &   238.33  $\pm$   1.53  & -2.73  &   2.00  &  -6.80  & 2.00    & 10.72 & 35.8 & -\\
J131729.6-462521 & 13h 17' 29.62" &   -46$^{\circ}$ 25' 21.3"   &   196.70  $\pm$  0.93   & 1.22   &  11.00   &  1.10   &  11.00   &  10.63 & 42.5 & -\\
J132045.6-445053 & 13h 20' 45.61" &   -44$^{\circ}$ 50' 52.9"   &   300.91  $\pm$   1.15  & 4.02  &   1.00   & -5.20   &  1.00    & 10.83 & 44.3 & -\\
J132209.4-481432 & 13h 22' 09.41" &   -48$^{\circ}$ 14' 31.7"   &   175.63  $\pm$  0.62   & 3.45   &   9.00  &  -0.30   &    9.00    &  8.79 & 54.4 & -\\
J132430.7-472427 & 13h 24' 30.74" &   -47$^{\circ}$ 24' 26.5"  &    229.33  $\pm$   0.75   & 13.44 &  11.00 &  1.70 &  11.00 &  8.92  &  41.5 & - \\
J132446.8-472449 & 13h 24' 46.76" &   -47$^{\circ}$ 24' 48.6"   &   234.78  $\pm$   1.10  & 2.87  & 11.00   & 4.10   &  11.00   &  10.51 & 42.0 & u\\
J132517.5-472427 & 13h 25' 17.53" &   -47$^{\circ}$ 24' 26.6"  &    228.57  $\pm$  0.57   & 1.82  &  3.00 & -5.10 & 3.00 &  9.54 & 57.1 & -\\
J132521.3-473654 & 13h 25' 21.32" &   -47$^{\circ}$ 36' 54.0"   &   239.65  $\pm$  0.40   &  7.32   &   3.00   &  -5.50    &   3.00    &  8.66  &  91.9 & -\\
J132521.8-452320 & 13h 25' 21.79" &   -45$^{\circ}$ 23' 19.8"   &   162.52  $\pm$  0.62   &  -1.86  &  12.00   &  11.20  &   12.00     & 10.45 & 36.2 & -\\
J132545.2-473238 & 13h 25' 45.15" &   -47$^{\circ}$ 32' 38.3"   &   217.80  $\pm$  0.68   & 2.46  &  3.00   &  -7.10   &   3.00    &  9.39 & 49.9 & v, u\\
J132551.2-472702 & 13h 25' 51.20" &   -47$^{\circ}$ 27' 01.8"   &   222.19  $\pm$  0.49   & 9.47   &   3.00   & -5.30   &   3.00    &  9.18 & 47.8 & v, u\\
J132552.0-473016 & 13h 25' 52.00" &   -47$^{\circ}$ 30' 16.3"  &    213.89  $\pm$   0.89   & -58.49  &  3.00 &  -5.50 &  3.00 &  8.81 &  52.9 & v, u\\
J132558.7-473610 & 13h 25' 58.72" &   -47$^{\circ}$ 36' 09.8"   &   235.57  $\pm$   2.20  & 7.01   &   3.00   &  -4.00   &   3.00   &   9.11 &  34.7 & -\\
J132601.7-474034 & 13h 26' 01.74" &   -47$^{\circ}$ 40' 33.6"   &   253.80  $\pm$  0.65   & 5.79   &   3.00  & -0.20    &   3.00   &   9.27 & 66.4 & v, u\\
J132609.1-472720 & 13h 26' 09.06" &   -47$^{\circ}$ 27' 19.5"  &    236.99  $\pm$  0.55   & 5.59  &  2.00 &  -9.40 &  2.00 &  9.66 & 51.2 & v, u\\
J132614.6-472123 & 13h 26' 14.56" &   -47$^{\circ}$ 21' 22.9"   &   244.65  $\pm$  0.60   & 11.38    &   3.00   &  -8.70   &    3.00    &  8.67 &  44.2 & v\\
J132623.7-474243 & 13h 26' 23.66" &   -47$^{\circ}$ 42' 42.5"   &   233.22  $\pm$  1.00  & -1.33   &   8.00   &  1.70  &    8.00   &   9.19 & 30.1 & -\\
J132629.6-473701 & 13h 26' 29.62" &   -47$^{\circ}$ 37' 01.4"   &   231.27  $\pm$   1.13  & 2.82   &   3.00   &  -9.90   &   3.00    &  9.35 & 59.0 & -\\
J132639.0-474359 & 13h 26' 38.96" &   -47$^{\circ}$ 43' 58.5"   &   226.20  $\pm$   1.22  & 0.16  &  3.00  &  -3.20    &  3.00  &  10.05 & 41.2 & -\\
J132639.3-472035 & 13h 26' 39.27" &   -47$^{\circ}$ 20' 34.8"   &   216.40  $\pm$  0.64   & -68.91  &  12.00   &  27.50   &   12.00    &  9.24 & 66.4 & v, u\\
J132646.2-471415 & 13h 26' 46.17" &   -47$^{\circ}$ 14' 15.2"   &   223.66  $\pm$  0.66   & 4.27    &   2.00   &  -4.80   &    2.00   &   9.29 & 63.2 & -\\
J132654.3-474605 & 13h 26' 54.34" &   -47$^{\circ}$ 46' 05.1"   &   240.49  $\pm$  0.64   & 2.56  &   3.00   &  -11.10    &   3.00    &  10.01 & 80.3 & -\\
J132704.4-443003 & 13h 27' 04.43" &   -44$^{\circ}$ 30' 03.1"   &   328.80  $\pm$   1.32  & -5.99  &  9.00    &  26.40   &   9.00  &  11.08 & 24.9 & v\\
J132709.6-472052 & 13h 27' 09.60" &   -47$^{\circ}$ 20' 51.5"  &    236.96  $\pm$  0.40   & 10.53 &  2.00 &  -6.80 &  2.00 &  9.05 & 60.6 & v, u\\
J132710.5-473701 & 13h 27' 10.54" &   -47$^{\circ}$ 37' 00.5"   &   248.80  $\pm$   1.11  & -89.77 & 18 &  -86.90 &  18.00 &  9.08 & 44.4 & v\\
J132726.0-473060 & 13h 27' 25.97" &   -47$^{\circ}$ 30' 59.9"   &   261.82  $\pm$   1.19  & 5.45   &   3.00   &  -8.30    &   3 .00  &   9.37 & 59.0 & v, u \\
J132753.7-472442 & 13h 27' 53.72" &   -47$^{\circ}$ 24' 42.1"   &   234.62  $\pm$   1.86  & -0.09  &  13.00  &  -15.50   &  13.00   &  10.61 & 38.3 & v, u\\
J132754.7-471932 & 13h 27' 54.71" &  -47$^{\circ}$ 19' 32.2"   &   248.04   $\pm$   2.23  & 6.47 &  2.00  &  -3.00  &  2.00 &  9.08 & 24.5 & -\\
J132757.3-473638 & 13h 27' 57.31" &   -47$^{\circ}$ 36' 38.1"   &   252.55  $\pm$  0.97   & 70.18  &  15.00   & 18.30   &  15.00  & 10.58 & 42.2 & v, u\\
J132800.8-473247 & 13h 28' 00.77" &   -47$^{\circ}$ 32' 47.0"   &   248.72  $\pm$  0.64   & -1.00  &  3.00    & -9.60    &   3.00    &  10.34 & 67.9 & v, u\\
J132804.8-474504 & 13h 28' 04.83" &  -47$^{\circ}$ 45' 04.1"   &    220.21  $\pm$  0.75   & 4.86 &  2.00  &  -9.60 &  2.00 &  8.65 & 34.3 & -\\
J132813.6-472424 & 13h 28' 13.57" &   -47$^{\circ}$ 24' 23.5"   &   231.31  $\pm$   1.04  & 3.74   &   9.00  &  -1.90   &   9.00   &   9.41 & 30.0 & -\\
J132815.0-473739 & 13h 28' 15.04" &   -47$^{\circ}$ 37' 39.4"   &   237.78  $\pm$   1.88  & 28.64  &  11.00  &  -8.00   &  11.00   &   9.94 & 20.0 & -\\
J132816.9-472956 & 13h 28' 16.90" &   -47$^{\circ}$ 29' 56.0"   &   245.12  $\pm$  0.56   & 8.18    &   3.00   &  -8.30   &   3.00    &  9.02 & 55.4 & -\\
J132833.8-473206 & 13h 28' 33.81" &   -47$^{\circ}$ 32' 05.6"   &   239.87  $\pm$  0.79   & -1.10   &  3.00   &  -9.90  &   3.00    &  8.84    &  64.7 & -\\
J132839.9-472633 & 13h 28' 39.93" &  -47$^{\circ}$ 26' 32.9"   &   228.40   $\pm$   2.89  & 5.51  &  3.00 &  -9.20 &  3.00  &  8.82  &  23.9 & v \\
J132918.9-471924 & 13h 29' 18.86" &   -47$^{\circ}$ 19' 23.9"   &   233.41  $\pm$  0.64   & 1.86  &  9.00 & -0.90 & 9.00 &  9.01 & 60.0 & -\\
J132936.8-500005 & 13h 29' 36.76" &  -50$^{\circ}$ 00' 05.4"   &   199.04   $\pm$  0.83    & -0.58 &  2.00 &      3.60 &    2.00 &  8.03  &  44.3 & -\\
J133106.0-483312 & 13h 31' 06.04" &   -48$^{\circ}$ 33' 11.9"   &   205.74  $\pm$  0.92   & 1.79  &  11.00   &  5.10  &  11.00   &  10.25 & 41.1 & -\\
J133257.2-492045 & 13h 32' 57.15" &   -49$^{\circ}$ 20' 45.2"   &   182.31  $\pm$  0.67   & 2.60  &  10.00  &   7.70  &  10.00   &  9.71 & 55.8 & v\\
J133328.7-441903 & 13h 33' 28.71" &   -44$^{\circ}$ 19' 02.8"  &    223.66  $\pm$   0.65   & 8.97 &  2.00 &  -7.10 &  2.00 &  7.73 & 74.8 & -\\
J133430.0-474615 & 13h 34' 29.96" &  -47$^{\circ}$ 46' 14.8"   &   161.65   $\pm$   1.49  & 5.74 &  3.00 &  -11.80  &  3.00  &  8.89  &   21.9 & - \\
J133536.3-450533 & 13h 35' 36.28" &  -45$^{\circ}$ 05' 33.0"   &   168.34   $\pm$  0.79   & -7.91 &  2.00 &  -2.00 &  2.00 &  8.41 &  58.7 & -\\
J133609.4-440854 & 13h 36' 09.39" &   -44$^{\circ}$ 08' 53.6"  &    181.41  $\pm$    1.30  & -7.99 &  1.00 &  -19.20  &  1.00  &  9.80 & 29.3 & -\\
J133628.5-461932 & 13h 36' 28.54" &   -46$^{\circ}$ 19' 32.1"  &    162.48  $\pm$  0.97   & 3.42 &  12.00 &  -1.60 &  12.00 &  9.45 & 30.7 & -\\
J133939.4-490014 & 13h 39' 39.41" &   -49$^{\circ}$ 00' 13.5"   &   236.59  $\pm$  0.56   & 0.79   &   2.00   &  -35.00   &    2.00    &  9.66 & 61.5 & -\\
J134508.6-492832 & 13h 45' 08.64" &  -49$^{\circ}$ 28' 32.1"   &    217.44  $\pm$   0.68  & -3.59 & 12.00  &  -11.10  &  12.00  &  8.52  &  68.8 & -\\
\hline 
\label{tab:NGC5139_cands}
\end{tabular}
\end{minipage}
\end{table*}

We found a total of 52 stars to be potential members of $\omega$ Cen based on their RVs (Table \ref{tab:NGC5139_cands}). 31 of those stars are inside the tidal radius and the proper motion diagram in Fig.\ref{fig:NGC5139_pos} shows that 45 stars share an apparent proper motion close to the value for $\omega$ Cen ($\mu_{\alpha}$ cos($\delta$) = -3.43,  $\mu_{\delta}$ = -3.57), suggesting these stars are probable members of $\omega$ Cen. 


\begin{table*}
 \centering
 \begin{minipage}{370mm}
  \caption{NGC 362 candidates selected from RAVE data and their parameters. The last column indicates the \emph{crowding}.}
  \begin{tabular}{@{}rrrrrrrrrrrrr@{}}
  \hline
  \hline
ID & R.A. & Decl. & RV & $\mu_{\alpha}$ cos($\delta$) & $\sigma_{\mu_{\alpha}}$ & $\mu_{\delta}$ & $\sigma_{\mu_{\delta}}$ & \emph{J} & S/N & C.  \\
   &&& (km/s) & (mas/yr) & (mas/yr) & (mas/yr) & (mas/yr) & \\
 \hline
J004905.3-733108  &  00h 49' 05.27'' & -73$^{\circ}$ 31' 07.9''  &   229.27 $\pm$  0.78 &  -6.59  &  2.00  &  -8.00 &  2.0 & 9.20 & 31.2 & -  \\
J004217.1-740615  &  00h 42' 17.14'' & -74$^{\circ}$ 06' 15.3''  &   266.93  $\pm$  0.79   &  5.01   &   1.00  &   -15.10 & 1.0 &  9.40 & 35.5 & -  \\
J005038.4-732818  &  00h 50' 38.39'' & -73$^{\circ}$ 28' 18.3''    &  221.15  $\pm$   7.19  &  0.29  &  1.00  &    -1.90  &  1.00  &  11.07 & 36.0 & -   \\  
J010313.6-705037  &  01h 03' 13.62'' & -70$^{\circ}$ 50' 36.8''   &  220.97  $\pm$  1.09  &  -   & - & - & - & - & 58.6 & v, u\\
J010314.7-705115  &  01h 03' 14.67'' & -70$^{\circ}$ 51' 15.3''    &  225.40  $\pm$   1.48  & -  & - & - & - & - &  26.4 & -\\
J010314.7-705059  &  01h 03' 14.74'' & -70$^{\circ}$ 50' 58.9''  &   225.01  $\pm$  0.55   &   - & - & - & - & - & 106.1 & v, u\\
J010315.1-705032  &  01h 03' 15.10'' & -70$^{\circ}$ 50' 32.3''  &   233.88  $\pm$  1.10  &   - & - & - & - & - & 40.6 & v, u\\
J010319.0-705051  &  01h 03' 19.03'' & -70$^{\circ}$ 50' 51.4''    &  222.78  $\pm$   1.01  &  -  & - & - & - & - & 51.4 & v, u\\
J010335.7-705052  &  01h 03' 35.71'' & -70$^{\circ}$ 50' 52.0''   &  228.24 $\pm$   0.74   &   7.94  & 2.2 & 58.6 & 2.2 & 9.73 & 38.6 & v, u \\
J011655.9-690607  &  01h 16' 55.87'' & -69$^{\circ}$ 06' 07.3''    &  236.51  $\pm$   1.80  &  11.80 &  10.00 &  -7.00 &   10.00  &  9.92 &  18.9 & -    \\
\hline 
\label{tab:NGC362_cands}
\end{tabular}
\end{minipage}
\end{table*}

\begin{figure}
  \centering  
  \includegraphics[width=1.02\columnwidth]{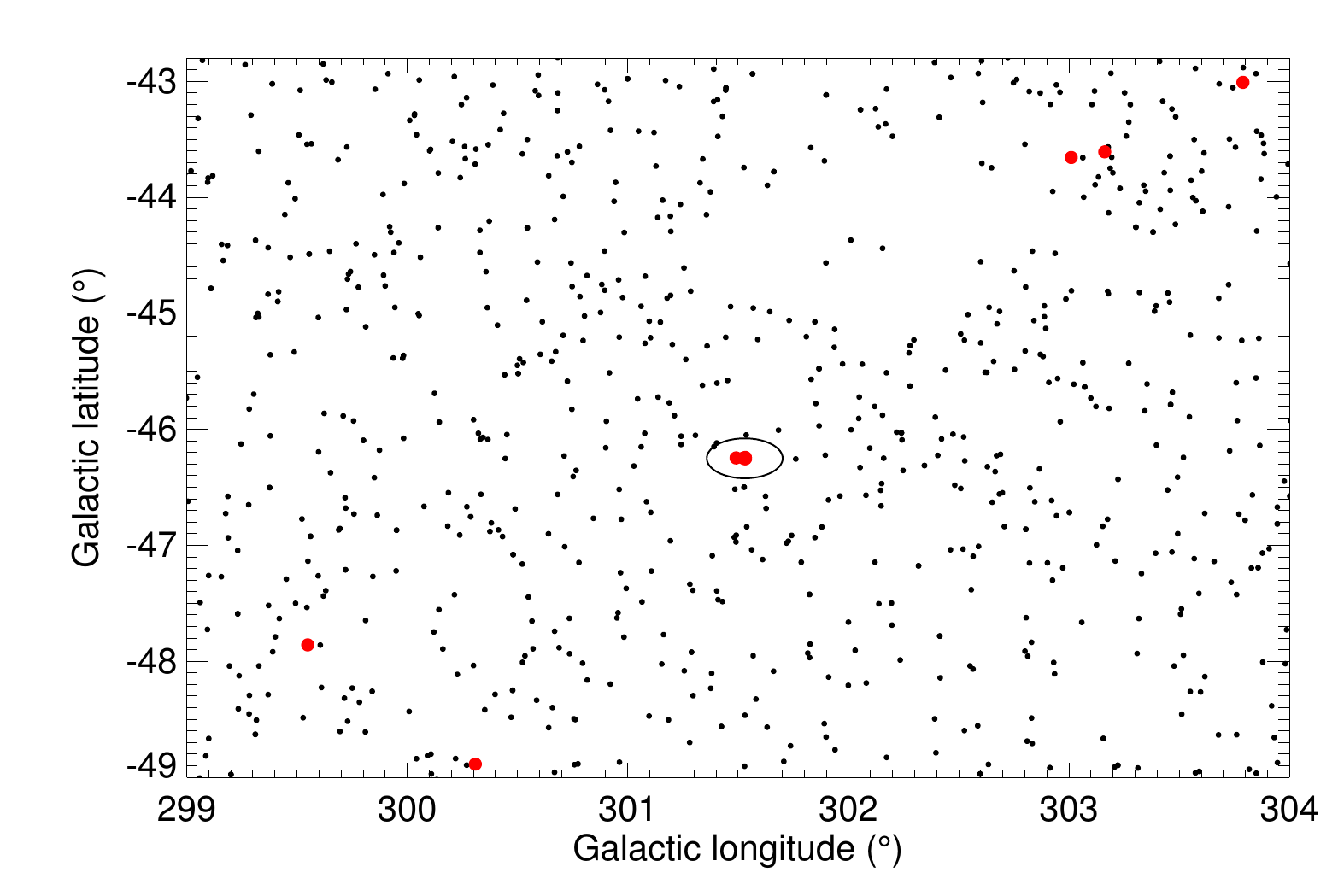}
  \includegraphics[width=1.02\columnwidth]{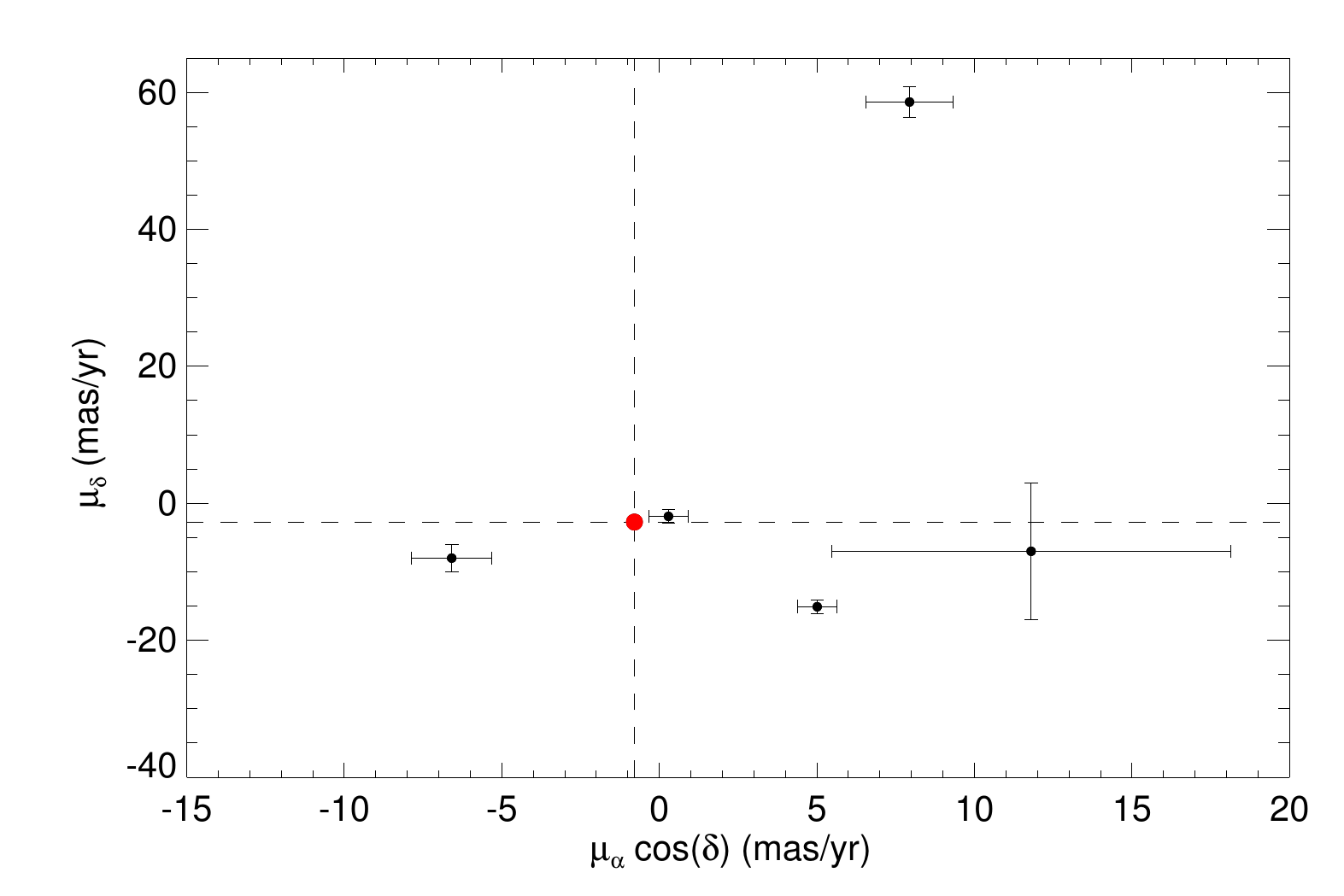}
  \caption{\emph{Top panel:} The positions (in Galactic coordinates) for the stars in the RAVE catalogue in the region of NGC 362. The red dots are potential members of NGC 362 selected based on RV. The black circle is the tidal radius according to the value given in \citet{1996AJ....112.1487H} (2010 edition) for this cluster. We identified a number of stars outside of the tidal radius as candidate members of the cluster. \emph{Bottom panel:} The proper motion plane (mas/yr) for the selected candidates. The red dot indicates the absolute proper motion of the cluster \citep{2007AJ....134..195C, 2010AJ....140.1282C, 2013AJ....146...33C}.}
  \label{fig:pos_NGC362}
\end{figure}

10 stars were selected using the RV for the cluster {NGC 362} (Table \ref{tab:NGC362_cands}). We find five stars inside the tidal radius (see Fig.~\ref{fig:pos_NGC362}). These stars are very close to each other but, unfortunately, we do not have proper motion information for them. The remaining stars are clearly outside of this radius. Fig.~\ref{fig:pos_NGC362} shows that the proper motion of these stars is large with respect to the cluster ($\mu_{\alpha}$ cos($\delta$) = -0.79,  $\mu_{\delta}$ = -2.73). There is one star, J005038.4-732818, with proper motions close to zero and a similar RV to the NGC 362, that might be associated with the cluster despite being outside of the tidal radius.

In summary, using the RVs and proper motions for stars in the RAVE catalogue and cluster tidal radii from the literature we were able to identify -- without any abundance selection bias --  
likely candidate members for several nearby galactic globular clusters.



\subsection{Crowding}

The 6dF spectrograph used for the RAVE survey placed wide (6.7 arcsec)  fibres on the sky; the size of these fibres could present problems because of the potential for overlapping background spectra of other stars, especially in the dense stellar regions around globular clusters.  Fibres in which the composite spectra of more than one star overlap should not be used for the calibration of stellar parameters, as the accuracy of the resulting analyses can be at best uncertain. To identify targets with possible crowding problems we have done two different tests. Stars classified as problematic with respect to general crowding or the presence of close neighbours via a visual inspection of DSS/2MASS finder charts are marked with ``v" in Tables~ \ref{tab:NGC3201_cands}, \ref{tab:NGC5139_cands} and \ref{tab:NGC362_cands}. We also used UCAC4 \citep{2013AJ....145...44Z} to mark suspicious cases; namely, we checked if there were any UCAC4 stars within 9 arcsec, or any relatively bright stars at larger separations. The stars for which this is true are marked with ``u" on the tables. In many cases, the crowding problem also led to uncertain (large) proper motions in the PPMXL catalogue, especially in the case of NGC 362.  

\section{Stellar parameters in the RAVE survey}

The RAVE survey uses the wavelength region $\lambda\lambda$8410 - 8795 \AA~ for the determination of the main stellar parameters in the atmosphere of the observed stars. In this region we find the Ca II triplet, iron and $\alpha$-element lines; these features are prominent spectral lines even for spectra with low signal-to-noise and for metal-poor stars, making this region useful for metallicity estimations over a broad range of stellar properties \citep{2001MNRAS.326..981C, 2011A&A...535A.106K}.

The methodology for determining stellar parameters in the survey has undergone several revisions. \citet{2008AJ....136..421Z} used a penalised $\chi^{2}$ method employing an extensive grid of synthetic spectra calculated from the latest version of Kurucz stellar atmosphere models for the first and second RAVE data releases. From comparison with external data sets, \citet{2008AJ....136..421Z}  estimated errors in stellar parameters for a RAVE spectrum with an average signal-to-noise ratio (S/N $\sim$ 40) to be 400 K in temperature, 0.5 dex in gravity, and 0.2 dex in metallicity. \citet{2011AJ....141..187S}, for the third data release, used new synthetic spectra for intermediate metallicities that were added in order to provide a more realistic spacing toward the densest region of the observed parameter space, and thereby remove biases toward low metallicity. They also used a  new continuum normalisation which signiÞcantly reduced the correlation between metallicity and S/N, masked bad pixels and improved the radial velocity zero-point. Finally, \citet{2013AJ....146..134K} computed the stellar atmospheric parameters using a new pipeline, based on the algorithms of MATISSE \citep{2006MNRAS.370..141R, 2011A&A...535A.106K} and DEGAS (DEcision tree alGorithm for AStrophysics) for the fourth RAVE data release (DR4). Spectral degeneracies and 2MASS photometric information are also taken into consideration. In this study we use the DR4 stellar parameters, as derived in \citet{2013AJ....146..134K}. \citet{2011AJ....142..193B} presented elemental abundances derived from RAVE spectra for the elements Mg, Al, Si, Ca, Ti, Fe, and Ni, through a processing pipeline in which the curve of growth of individual lines is obtained from a library of absorption line equivalent widths to construct a model spectrum, that is then matched to the observed spectrum via a $\chi^{2}$-minimisation technique. 

In this section we use a subset of the candidate cluster members identified in the previous section, namely those stars which match the radial velocities and fall within the tidal radii of the three globular clusters discussed above -- NGC 3201, $\omega$ Cen and NGC 362 --  to test the validity of the stellar parameters derived from RAVE spectra. Since most GGCs appear to have ages $>$ 10 Gyr (e.g. \citealt{2010ApJ...708..698D, 2013ApJ...775..134V}) we decided to use isochrones for the ages determined for these clusters in the literature to test T$_{\rm eff}$, [m/H] and $\log$ g determinations. Although $\omega$ Cen may contain younger stellar populations (e.g., \citealt{2007ApJ...663..296V}), it is very difficult to derive the ages of single giant stars to within a few Gyr given that temperatures along computed red giant branches (RGBs) are very sensitive to many aspects of stellar physics (notably convection) and because the location of the RGB is much less dependent on age than on metal abundances (e.g., \citealt{2012ApJ...755...15V}). We also explore the abundances for the elements Mg, Al, Si, Ca, Ti, Fe, and Ni for the potential members of the clusters.


\subsection{The case of NGC 3201}

NGC 3201 shows very peculiar kinematic characteristics. The cluster has an extreme radial velocity and a highly retrograde orbit \citep{2007AJ....134..195C}. Thus kinematically NGC 3201 appears likely to be of extragalactic origin; however \citet{2013MNRAS.433.2006M} claim that its chemical evolution was similar to most other, presumably ``native'', GGCs. NGC 3201 is a low mass halo cluster and the existence of star-to-star metallicity variations remains controversial. In agreement with the findings of \citet{1981ApJ...248..612D}, \citet{2009A&A...508..695C} found no significant variations of [Fe/H] in NGC 3201 in their analysis of high-resolution spectra of hundreds of stars, and \citet{2013MNRAS.433.2006M} similarly found no evidence for any intrinsic Fe abundance spread, except for one star in their sample. On the other hand, \citet{1998AJ....116..765G} and \citet{2013ApJ...764L...7S} identified a spread in [Fe/H] in cluster stars at least as large as 0.4 dex, even though \citet{2003PASP..115..819C} were not able to confirm the presence of a significant spread in metallicity within the cluster greater than about 0.3 dex. Iron abundance variations have been found in the most massive globular clusters (e.g., M22, M54, $\omega$ Cen), suggesting multiple star formation episodes and metal enrichment via Type Ia supernova events. \citet{2013ApJ...764L...7S} concluded that a real [Fe/H] spread, if it did exist, would support the idea that NGC 3201 was initially far more massive, formed outside the Milky Way, and was subsequently captured. Recently, a possible solution to the controversy surrounding the  metallicity spread in NGC 3201 has been proposed by \citet{2015ApJ...801...69M}, who demonstrated that the metal-poor component claimed by \citet{2013ApJ...764L...7S} is composed of asymptotic giant branch stars that could be affected by non-local thermodynamical equilibrium effects driven by iron overionisation, and therefore concluded that there is no evidence of intrinsic iron spread.


\begin{figure*}
  \centering  
  \includegraphics[width=0.8\linewidth]{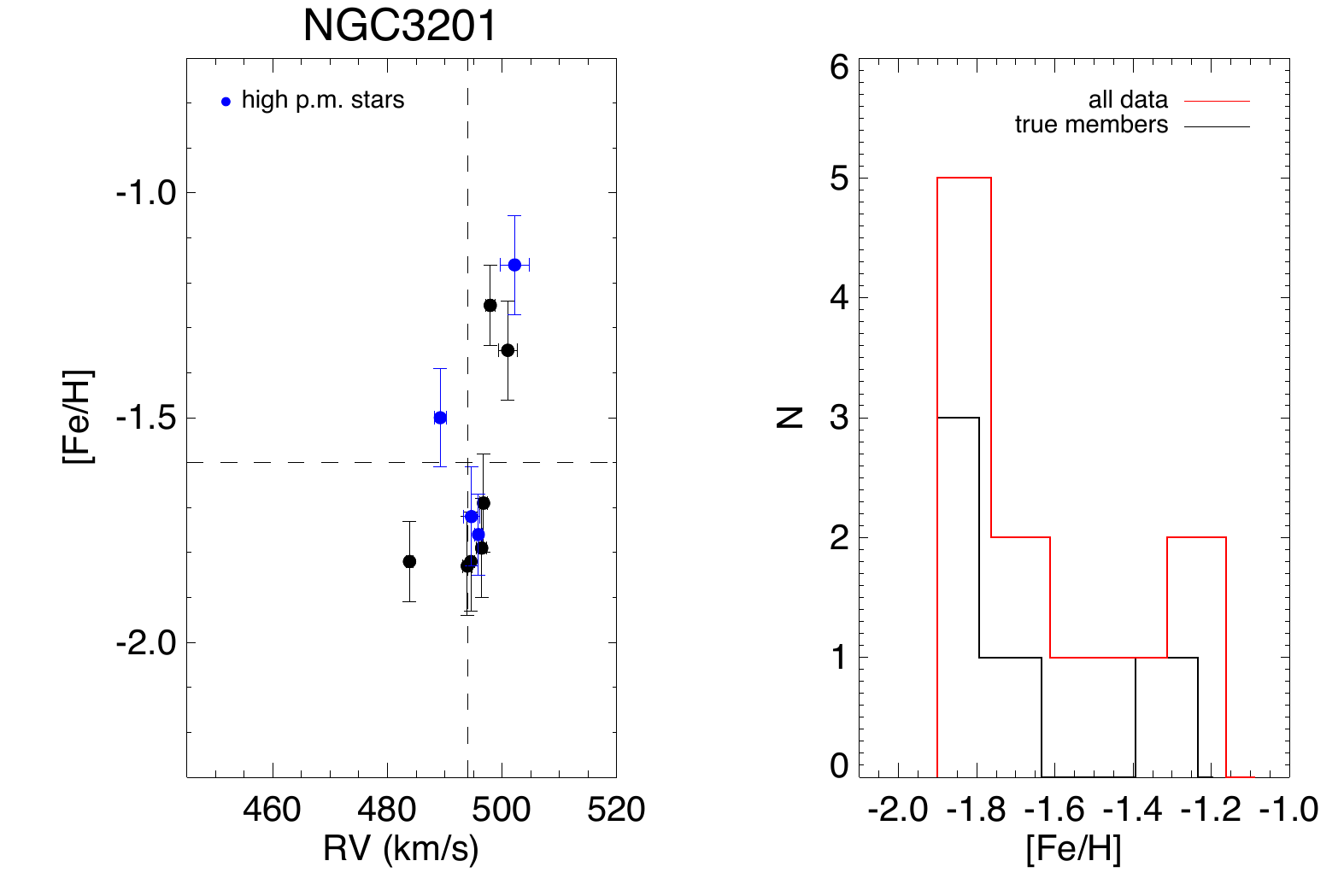}
  \caption{\emph{Left panel:} Radial velocity-abundance plot for RV candidate members which fall within the tidal radius of NGC 3201. The dashed lines indicate the nominal RV and [Fe/H] for the cluster reported in the literature. The blue dots are stars with large proper motions with respect to the measured proper motion for the cluster. \emph{Right panel:} The metallicity distribution function for all RV candidate members which fall within the tidal radius (in red) and for those which also have proper motion measurements consistent with that of the cluster (black line).}
  \label{fig:RV_meta_NGC3201}
\end{figure*} 
 
\subsubsection{NGC 3201 abundances, temperatures and gravities}

RAVE DR4 metallicities were calibrated using [Fe/H] from the literature and dedicated observations of calibration stars. Since the metallicity measurement is dominated by the Ca II lines, we have the overall metallicity [M/H]$_{DR4}$ $\approx$ [Fe/H] + a small correction from $\alpha$-elements. In this work we use [M/H]$_{DR4}$ = [Fe/H], see Fig.~8 in \citet{2013AJ....146..134K}. For this cluster we adopt $<$[Fe/H]$>$ = -1.5 $\pm$ 0.1 dex \citep{2009A&A...508..695C}, which agrees well with other determinations (e.g.  \citealt{1998AJ....116..765G, 2013MNRAS.433.2006M})


From the highest likelihood members of NGC 3201 -- i.e., those with RVs and proper motions consistent with membership and which also fall within the tidal radius -- we find two groups of stars (see Fig.~\ref{fig:RV_meta_NGC3201}, black dots). The largest group has a mean [Fe/H] of -1.80 $\pm$ 0.11 dex while the second group contains only two stars and has a $<$[Fe/H]$>$ $\sim$ -1.30 $\pm$ 0.10 dex. The RAVE SPP is able to identify these stars as metal-poor and the mean [Fe/H] value for this cluster from the RAVE pipeline is -1.55 $\pm$ 0.10, in excellent agreement with high-resolution spectroscopy as indicated above. The first group has an observed scatter of 0.05 dex, a lower value than the nominal uncertainty of the metallicity measurement, suggesting that there is no evidence for an intrinsic scatter in the metallicity in terms of RAVE errors. The second group exhibits an observed scatter of 0.07 dex, and again there is no evidence for an intrinsic scatter, although there are only two stars in the group. Thus we have found two clearly different metallicity groups that are potentially members of NGC 3201 based solely on their kinematics and proper motions. However these two groups could be just an artefact of selection effects on our small sample size, as other authors with larger data-sets have not reported such a bimodal distribution. Combining the two groups we find an observed scatter of 0.24 dex for NGC 3201, larger than the expected metallicity uncertainties. RAVE metallicities suggest that an intrinsic scatter is observed in this cluster with a star-to-star metallicity variation from -1.25 to -1.83 dex, however the metallicity of the two metal-rich stars should be verified before we draw conclusions on the intrinsic scatter in this cluster. This spread in metallicity is in good agreement with \citet{1998AJ....116..765G} and \citet{2013ApJ...764L...7S}; however \citet{2013ApJ...764L...7S} found their stars ranged between -1.80 $<$ [Fe/H] $<$ -1.40, and \citet{1998AJ....116..765G} found a spread over -1.65 $<$ [Fe/H] $<$ -1.15. As mentioned above, some authors did not find a spread in metallicity in this cluster. Moreover, the published colour-magnitude diagrams for the RGB of NGC 3201 do not appear to show enough scatter to accommodate a range in [Fe/H] between -1.3 and -1.8 dex, in particular the CMDs corrected for differential reddening \citep{2001AJ....121.1522V}.

There are four stars with proper motions significantly different from the nominal proper motion for the cluster. These stars are shown in blue in Fig.~\ref{fig:RV_meta_NGC3201}. Two stars lie in the most metal-poor group, in good agreement with the overall metallicity; one star is in the metal-rich group but it is slightly more metal-rich; and one star has an intermediate metallicity. Could these stars with high proper motions in fact be halo stars and not members of NGC 3201? This scenario seems rather unlikely. There are not many stars with extreme heliocentric RVs ($>$ 400 km s$^{-1}$) in the Galactic halo \citep[e.g.,][although it should be noted that the RVs in these two studies are Galactocentric and not heliocentric]{2007MNRAS.379..755S, 2014A&A...562A..91P}.  For NGC 3201, RV$_{Gal}$ $\sim$ +275 km s$^{-1}$; this velocity is still high, but both papers identified objects in this velocity regime in the RAVE catalogue. These stars also have a metallicity range in good agreement with the abundances reported for this cluster, and they are at projected distances close to the cluster. However, typical halo stars can have metallicities in this range. Other potential explanations for the proper motion discrepancies are that these stars could be members of NGC 3201 but significantly closer to us than the main body of the cluster itself, or that the proper motions for these objects are simply imprecise or erroneous as these objects are in crowded fields. 

\citet{1996AJ....112.1487H} reported that the stars in NGC 3201 are around 10.5 Gyr old. \citet{2013ApJ...764L...7S} found that 14 Gyr isochrones with [$\alpha$/Fe]  = 0.0 fit the stellar parameters of the observed stars but they concluded that a younger age for the metal-poor stars would improve the isochrone fit. \citet{2010ApJ...708..698D} derived an age of 12.0 $\pm$ 0.75 Gyr from a deep HST CMD of NGC 3201, assuming that the stars in the cluster are $\alpha$-enhanced. Recently, \citet{2013MNRAS.433.2006M} reported an age of 11.4 Gyr, taking into account the (C+N+O) abundance in the isochrone age.

\begin{figure*}
  \centering  
   \includegraphics[width=0.80\linewidth]{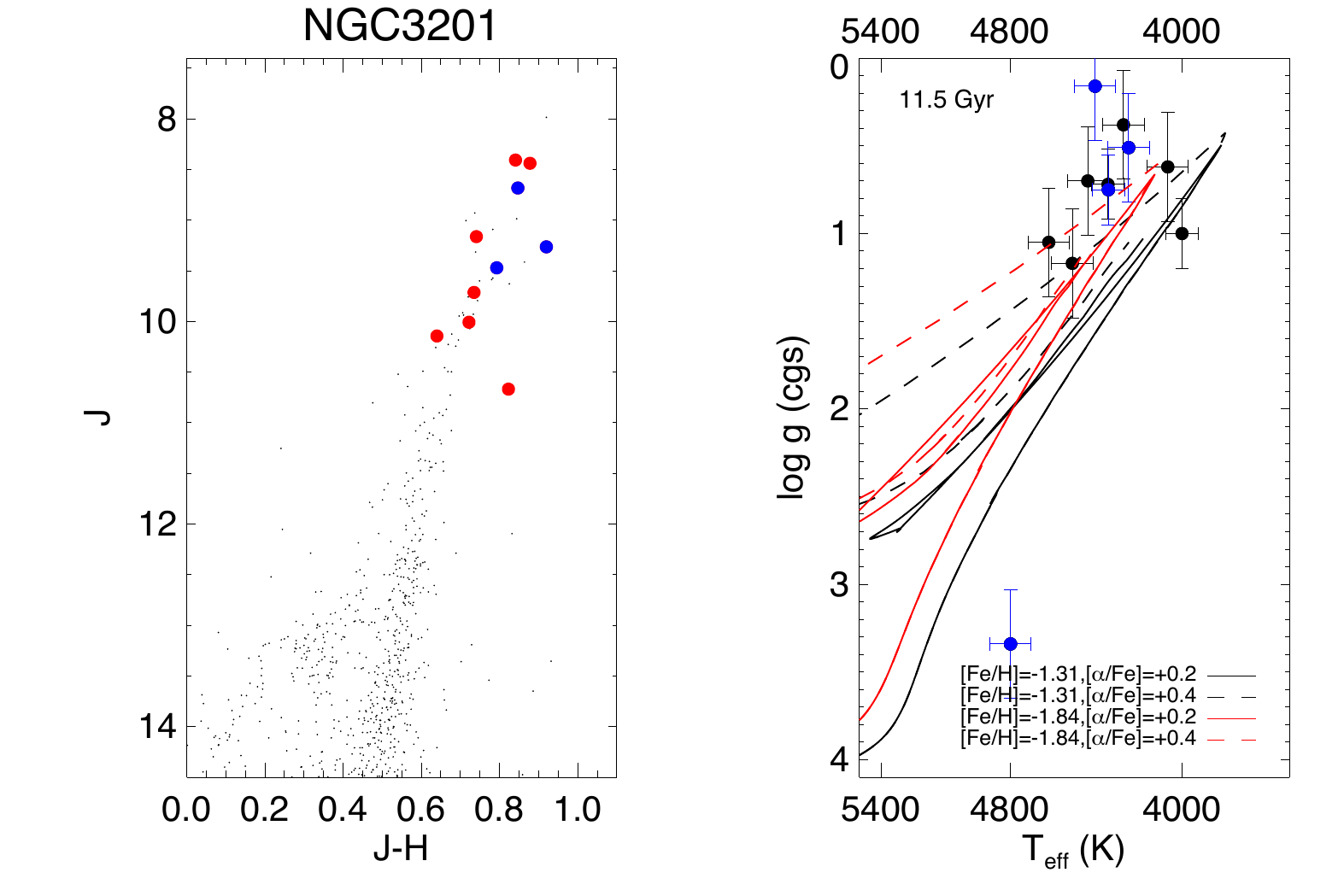}
  \caption{\emph{Left panel:} 2MASS J, J-H Colour-Magnitude Diagram (CMD) centred on NGC 3201, with a radius of 0.1 degrees. The large dots are the stars selected as being probable members of the cluster based on RV and location within the tidal radius; stars with large proper motions relative to the cluster are in blue. Note that most of the stars lie on the RGB of NGC 3201. \emph{Right panel:} Temperature-surface gravity diagram for the stars selected as probable members of NGC 3201, overplotted with BASTI isochrones for an age of 11.5 Gyr, [Fe/H] =  -1.3 and -1.8, and [$\alpha$/Fe] = +0.2 and +0.4 dex. The blue dots are stars with large proper motions.}
\label{fig:NGC3201_CMD}
\end{figure*} 

In addition, \citet{2013MNRAS.433.2006M} found, using only Mg, Si, Ca and Ti, a mean $\alpha$-element abundance for stars in NGC 3201 of +0.30 $\pm$ 0.06. This value is in good agreement with the [$\alpha$/Fe] values estimated by \citet{1998AJ....116..765G} and \citet{2009A&A...508..695C}. In order to test the effective stellar temperatures and the surface gravities derived from the RAVE spectra we compared them to BASTI isochrones \citep{2006MmSAI..77...71C}. We selected isochrones with an age = 11.5 Gyr and [Fe/H] = -1.31 and -1.84, the range in metallicity we found for this cluster. We also made use of $\alpha$-enhanced isochrones, in particular, [$\alpha$/Fe] = +0.2 and +0.4. Fig.~\ref{fig:NGC3201_CMD} shows the Colour-Magnitude Diagram (CMD) using 2MASS J and H bands. From the 2MASS catalogue \citep{2003tmc..book.....C} we selected stars inside a circle with a radius of 0.1 degrees centred on NGC 3201. The CMD shows the RGB, AGB and the HB of NGC 3201. We find that the selected stars from the RAVE catalogue lie in the upper part of the RGB indicated by red dots in Fig.~\ref{fig:NGC3201_CMD}; the blue dots are stars with high proper motion with respect to the nominal value of the cluster as indicated above. On the right panel in Fig.~\ref{fig:NGC3201_CMD} we have the T$_{\rm eff}$ - $\log$ g diagram for the stellar members of NGC 3201 together with the BASTI isochrones. Generally there is a good fit within the errors for the given isochrones. However there is a group of stars for which, in the metallicity range reported here, a younger isochrone ($\sim$ 10.5 Gyr) would provide a better fit. If we accept the age of NGC 3201 as 11.4 Gyr and [$\alpha$/Fe] $\sim$ 0.3 dex \citep{2013MNRAS.433.2006M}, we can conclude from this exercise that the gravities are generally underestimated with respect to the isochrones. 

Note that there is one star with $\log$ g = 3.3, clearly outside the main group of stars and the isochrones. This star is the most metal-rich ([Fe/H] = -1.16) and has a high proper motion. The star is marked as potentially affected by crowding problems, and a close look at the spectrum shows that the ``star'' is either two separate stars in the same fibre or a spectroscopic binary (SB2). This could easily lead to an erroneous measurement of $\log$ g. Interestingly, however, its estimated metallicity is similar to the other two high metallicity stars, neither of which appear to be impacted by crowding problems or show composite spectra. 

In the \citet{2011AJ....142..193B} chemical catalogue we find five candidate stars with [$\alpha$/Fe] measurements (see Table~\ref{tab:NGC3201_stelpars}). We exclude the one star with [$\alpha$/Fe] = +0.64, as this is the object mentioned above, with a high surface gravity and a possible composite spectrum. For the other four stars, the measured $<$[$\alpha$/Fe]$>$ = +0.22 is in good agreement with high resolution studies, and we find $\sigma_{[\alpha/Fe]}$ = 0.18 dex. 

\begin{table*}
 \centering
 \begin{minipage}{370mm}
  \caption{NGC 3201 candidate members selected from RAVE data and their stellar parameters}
  \begin{tabular}{@{}rrrrrrrrrrrrr@{}}
  \hline
  \hline
ID & T$_{eff}$ (K) & [Fe/H] & $\log$ g (cgs) & [$\alpha$/Fe] & J-H    \\
     \\
 \hline
J101405.6-462841 &  4510 $\pm$ 96 & -1.83 $\pm$ 0.11 & 1.17 $\pm$ 0.31 & +0.19 & 0.72\\
J101640.5-463221 &  4619 $\pm$ 96 & -1.83 $\pm$ 0.11 & 1.05 $\pm$ 0.31 & +0.31 & 0.64\\
J101648.9-461807 &  4438 $\pm$ 96 & -1.79 $\pm$ 0.11 & 0.70 $\pm$ 0.31 & - & 0.74 \\
J101716.2-462533 &  4797 $\pm$ 96 & -1.16 $\pm$ 0.11 & 3.34 $\pm$ 0.31 & +0.64 & 0.91\\
J101725.9-462621 &  4248 $\pm$ 96 & -1.50 $\pm$ 0.11 & 0.51 $\pm$ 0.31 & - & 0.85 \\
J101731.6-462901 &  4272 $\pm$ 96 & -1.35 $\pm$ 0.11 & 0.38 $\pm$ 0.31 & - & 0.82  \\
J101738.6-462716 &  4404 $\pm$ 96 & -1.72 $\pm$ 0.11 & 0.16 $\pm$ 0.31 & - & -\\
J101751.5-462210 &  4342 $\pm$ 75 & -1.76 $\pm$ 0.09 & 0.75 $\pm$ 0.20 & - & 0.79\\ 
J101752.1-461407 &  4344 $\pm$ 75 & -1.82 $\pm$ 0.09 & 0.72 $\pm$ 0.20 & +0.40 & 0.73\\ 
J101859.1-463438 &  4000 $\pm$ 75 & -1.25 $\pm$ 0.09 & 1.00 $\pm$ 0.20 & - & 0.88   \\
J102025.9-464406 &  4066 $\pm$ 96 & -1.69 $\pm$ 0.11 & 0.62 $\pm$ 0.31 & -0.03 & 0.84\\
\hline
\label{tab:NGC3201_stelpars}
\end{tabular}
\end{minipage}
\end{table*}
  
 \subsubsection{NGC 3201 stellar distances}

To date there have been four studies \citep{2010A&A...511A..90B, 2010A&A...522A..54Z, 2010MNRAS.407..339B, 2013MNRAS.tmp.2584B} which address the challenge of calculating distances for the RAVE stars. These have primarily used atmospheric stellar parameters derived from the spectra, photometric colours of the stars and stellar evolutionary tracks, i.e., the method of spectrophotometric parallaxes. In this work we test the distances derived in \citet{2010A&A...522A..54Z} and \citet{2013MNRAS.tmp.2584B} using the stellar parameters from RAVE DR4 \citep{2013AJ....146..134K}. \citet{2010A&A...522A..54Z} assumed that stars follow standard stellar evolution tracks, as reflected by theoretical isochrones. They also assumed that interstellar reddening is negligible because the vast majority of RAVE stars lie at high Galactic latitudes ($\mid$ b $\mid$ $ >$ 20$^{\circ}$). With these assumptions, \citet{2010A&A...522A..54Z} determined the probability distribution function (PDF) for the absolute magnitude. The authors concluded that their derived distances of both dwarfs and giants match the astrometric distances of Hipparcos stars \citep{2007ASSL} to within $\sim$ 21$\%$ using the RAVE DR3 stellar parameters \citep{2008AJ....136..421Z}. Another approach, with some assumptions via \emph{prior functions}, is the Bayesian framework developed in \citet{2010MNRAS.407..339B} and \citet{2013MNRAS.tmp.2584B}. Prior functions play a key role in these new distances and they reflect the state of our knowledge of the Galaxy, with different prior probabilities based on models of the density of the three components of the Galaxy (thin disc, thick disc and halo). \citet{2013MNRAS.tmp.2584B} included the effects of interstellar dust by applying a prior that reflects increasing extinction with distance and higher extinction towards the Galactic plane using the Schlegel maps \citep{1998ApJ...500..525S} in the prior. The distance determination of \citet{2013MNRAS.tmp.2584B} made use of stellar isochrones which only went down to $\sim$ -0.9 dex in metallicity, thus neglecting more metal-poor stars. They concluded that the expectation of parallax may be the most reliable distance indicator and found a good agreement between the expectation values of the parallaxes and the values measured by Hipparcos for the very few stars in common, especially in the case of hot dwarfs.


\begin{figure*}
  \centering  
   \includegraphics[width=0.80\linewidth]{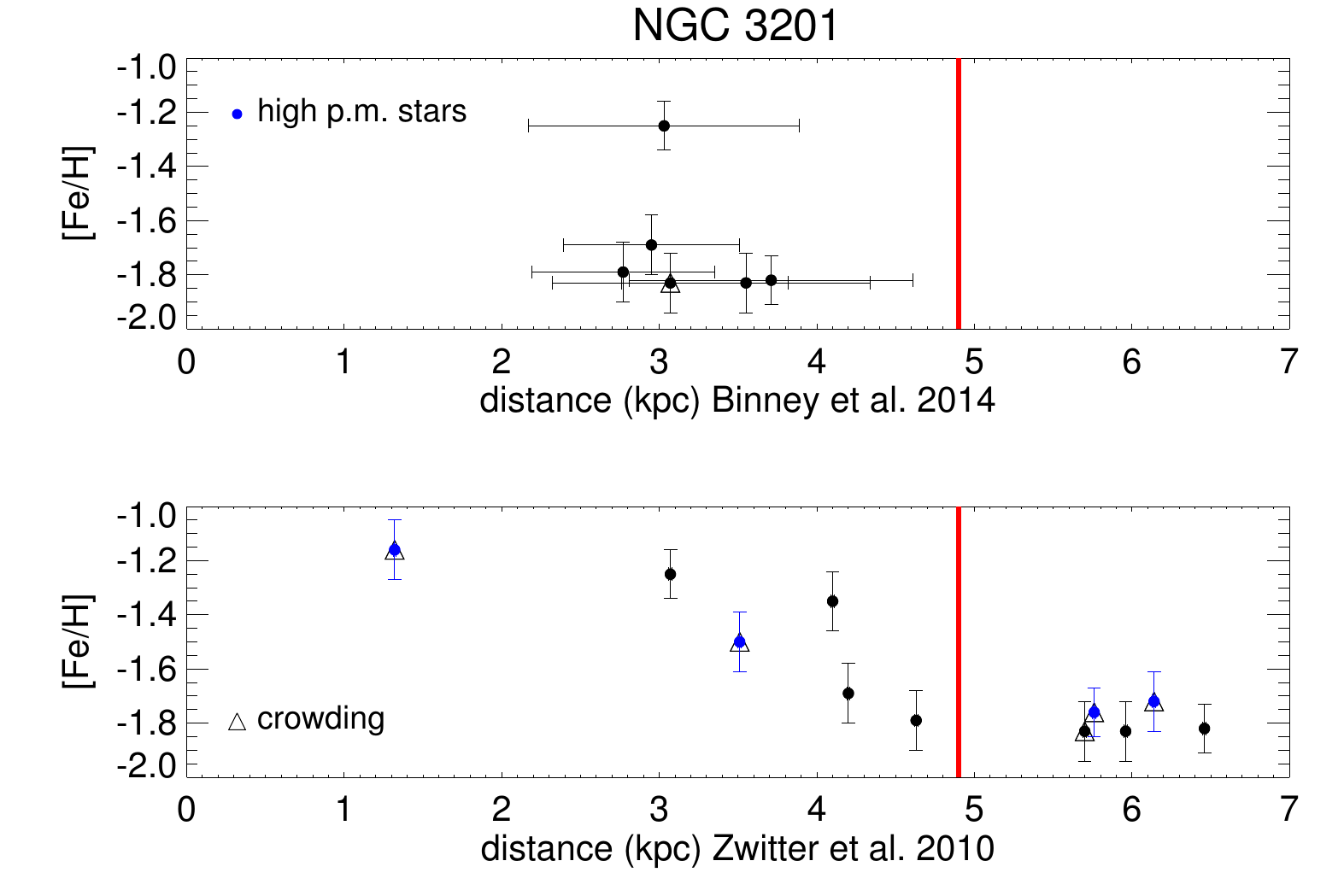}
  \caption{Distance vs. [Fe/H] for the RV and tidal-radius-selected candidate members of NGC 3201. In the top panel the distances are those derived in \citet{2013MNRAS.tmp.2584B}, in the bottom panel the distances are from \citet{2010A&A...522A..54Z}. Blue dots are stars with a large proper motion while triangles indicate stars affected by crowding. The vertical red line indicates the distance of the cluster given in \citet{1996AJ....112.1487H} (2010 edition).}
  \label{fig:NGC3201_dist}
\end{figure*} 

In Fig.~\ref{fig:NGC3201_dist} we show the estimated stellar distances for RV and tidal-radius-selected members of NGC 3201 together with their measured [Fe/H]. The top plot shows the distances from \citet{2013MNRAS.tmp.2584B} and the bottom shows the distances calculated by \citet{2010A&A...522A..54Z}; \citet{2013MNRAS.tmp.2584B} estimated the distances for six of the stars that we selected as likely members of the cluster, while in \citet{2010A&A...522A..54Z} we find estimated distances for 11 such objects. The blue dots in the figure are stars with high proper motions with respect to the nominal value for the cluster. For this study we adopted the true distance modulus from (\citealt{1996AJ....112.1487H}, see also \citealt{2003AJ....125..208L}), (m-M)$_{0}$ = 13.45, i.e., 4.9 kpc (red vertical line in Fig.~\ref{fig:NGC3201_dist}). \citet{2013MNRAS.tmp.2584B} distances for the six members have a small range from 2.7 to 3.7 kpc with a mean value of 3.1 kpc. If these objects are current members of the cluster and not a group of stars disrupted from NGC 3201, our results would suggest that the distances for the giant stars we identify as likely members of NGC 3201 are underestimated by $\sim$ 2 kpc. Note that in Fig.~\ref{fig:NGC3201_CMD} we found that the gravities for these stars appear to be slightly underestimated with respect to the selected isochrones using the latest estimations of age, [Fe/H] and [$\alpha$/Fe] for this cluster. Underestimated surface gravities could potentially affect the distance determination of the RAVE giants, although from a simplistic perspective we would expect lower surface gravities at the same temperature to yield higher intrinsic luminosities, and hence distances which are too \emph{large}, rather than too small. 



\citet{2010A&A...522A..54Z} calculated the distances for 11 members of the cluster (bottom panel in Fig.~\ref{fig:NGC3201_dist}). The blue dots in the figure are stars with high proper motion values with respect to the cluster. The distances found by \citet{2010A&A...522A..54Z} for these stars range from 1.3 to 6.5 kpc. They found a group of stars around 6 kpc, but two of those have a high proper motion. The reddening to NGC 3201 is rather larger, E(B-V)=0.24 mag \citep{1996AJ....112.1487H}, which may contribute to the distance spread, as \citet{2010A&A...522A..54Z} did not take reddening into account in their analysis. A comparison of the results of the two distance determination methods reveals that \citet{2013MNRAS.tmp.2584B} found their sample of stars to cover a small range in distances while the stars of \citet{2010A&A...522A..54Z} have a significant scatter in distance. 


Recently, a new distance calibration has been applied to these stars by P. McMillan, using the methodology of \citet{2013MNRAS.tmp.2584B} but incorporating isochrones with metallicities extending to lower than [Fe/H] = -0.9. The results of this work are consistent with the published distance for this cluster, with a mean distance of 4.2 $\pm$ 0.8 kpc for the members identified in this study. 

\subsection{The case of NGC 5139 ($\omega$ Centauri)}

The putative globular cluster $\omega$ Cen is a complex stellar system. If it is a globular cluster, it is the most massive such cluster in the Galaxy, in which clear evidence of multiple stellar populations has been detected \citep{1999Natur.402...55L, 2000AJ....119.1239S, 2010ApJ...722.1373J}. On the other hand, there is also evidence that $\omega$ Cen could be the stripped core of a dwarf elliptical galaxy \citep{2003MNRAS.346L..11B, 2006ApJ...637L.109B}. The very bound retrograde orbit supports the idea that the cluster entered the Galaxy as part of a more massive system whose orbit decayed through dynamical friction \citep{2002ARA&A..40..487F}. This cluster exhibits large star-to-star metallicity variations ($\sim$ 1.4 dex). Several studies have found [Fe/H] ranges from $\sim$ -2.1 to $\sim$ -0.7 dex using high-resolution spectroscopy for individual red giants in the cluster; moreover distinct peaks in the iron abundance distribution have been detected, suggesting different star formation episodes  \citep{2005ApJ...634..332S, 2007ApJ...663..296V, 2010IAUS..268..183M}. Estimates of the time periods spanned by these different star formation episodes vary widely, however. \citet{2007ApJ...663..296V, 2014ApJ...791..107V} found a spread in age of 5 Gyr between the youngest and oldest members of the cluster, ranging from 8 to 13 Gyr. In contrast, \citet{2004A&A...422L...9H, 2006ApJ...647.1075S} found an age spread on the order of 2 - 3 Gyr, and \citet{2005ApJ...634..332S} reported a small or negligible age dispersion. It is important to note that \citet{2012ApJ...746...14M} demonstrated a significant variation in the C+N+O content among $\omega$ Cen's stellar populations, which could easily have an impact on these age estimates.

We selected our highest probability $\omega$ Cen members from the RAVE catalogue using a combination of RVs, proper motions and location within the tidal radius. In the next section we explore the main stellar parameters for these objects.  

\subsubsection{Abundances, temperatures and gravities}

\begin{figure*}
  \centering  
   \includegraphics[width=0.80\linewidth]{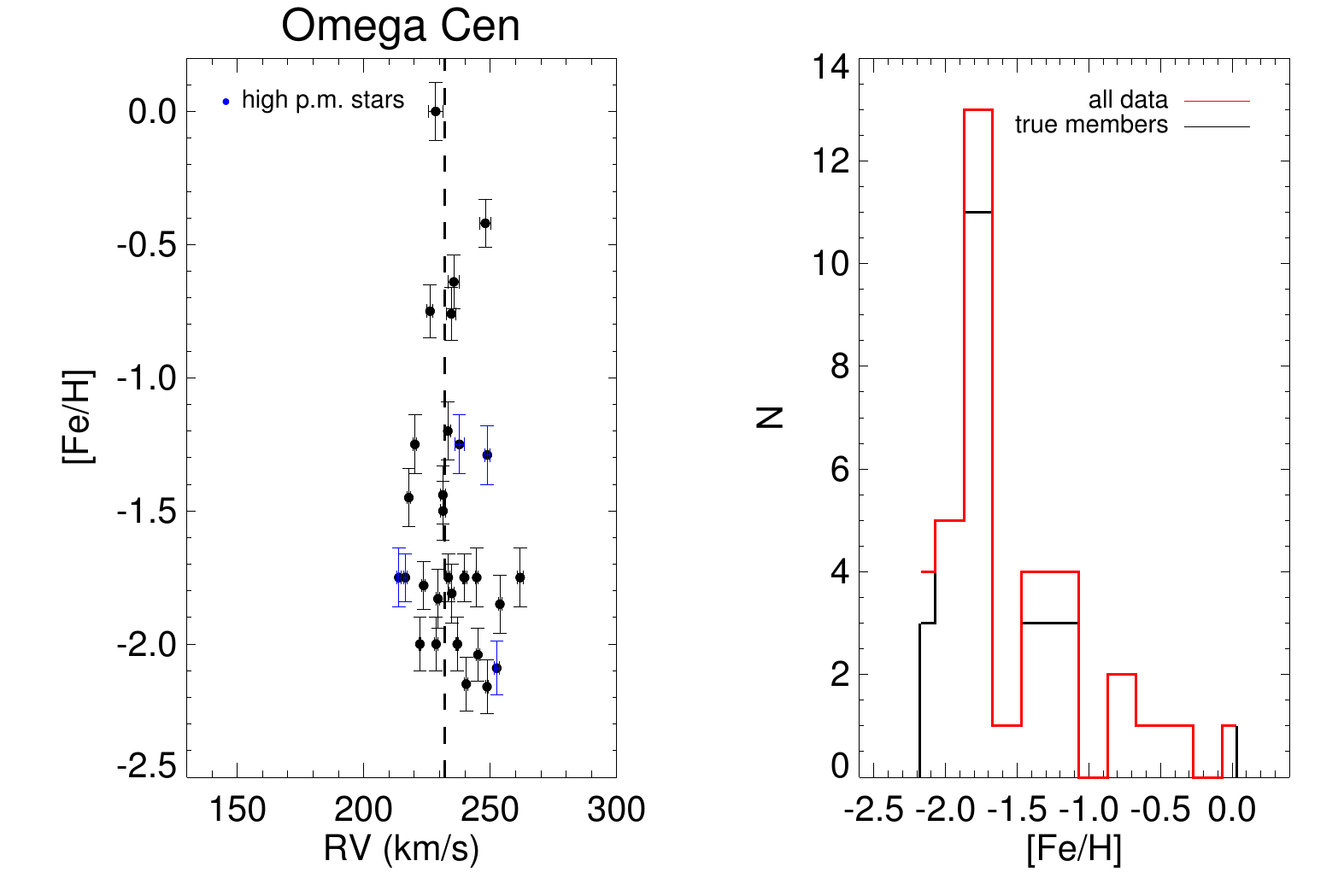}
  \caption{\emph{Left panel:} Radial velocity-abundance plot for RV candidate members of NGC 5139 ($\omega$ Cen). The dashed line indicates the nominal RV for the cluster reported in the literature. The blue dots are stars with large proper motions with respect to the measured proper motion for the cluster. \emph{Right panel:} Metallicity distribution function for all RV candidate members which fall within the tidal radius (in red) and for those which also have proper motion measurements consistent with that of the cluster (black line).}
  \label{fig:NGC5102_meta}
\end{figure*}

We find a large spread in [Fe/H] ($\sim$ 2.0 dex) for the selected candidate members of the cluster (see Fig.\ref{fig:NGC5102_meta}). The abundances range from approximately solar values ([Fe/H] $\sim$ 0.0) to metal-poor abundances ([Fe/H] $\sim$ -2.2). The blue dots in Fig.\ref{fig:NGC5102_meta} represent stars with RVs similar to the cluster and which lie inside the tidal radius, but which have large proper motions with respect to $\omega$ Cen. The distribution of [Fe/H] shows three peaks, at [Fe/H] $\sim$ -0.6, -1.3  and, the largest, at -1.8 (Fig.\ref{fig:NGC5102_meta}, right panel). The large range of metallicities observed in the RAVE targets is in very good agreement with high resolution spectroscopic studies as indicated above. The RAVE metallicity distribution function we obtain is also consistent with several star formation episodes in the history of the cluster. 

\begin{table*}
 \centering
 \begin{minipage}{370mm}
  \caption{NGC 5139 candidates selected from  RAVE data and their stellar parameters}
  \begin{tabular}{@{}rrrrrrrrrrrrr@{}}
  \hline
  \hline
ID & T$_{eff}$ (K) & [Fe/H] & $\log$ g (cgs) & [$\alpha$/Fe] & J-H    \\
     \\
 \hline
J131340.4-484714&        4516$\pm$          96& -1.36$\pm$  0.11&  0.83$\pm$  0.31&-  &0.74\\
J131602.3-480507&        4250$\pm$          86& -0.50$\pm$  0.10&  2.00$\pm$  0.20&-  &0.95\\
J131613.1-452004&        5201$\pm$         103& -0.84$\pm$  0.10&  3.20$\pm$  0.23&-  &0.74\\
J132430.7-472427&        4350$\pm$          96& -1.83$\pm$  0.11&  0.10$\pm$  0.31& +0.10  &0.48\\
J132446.8-472449&        4590$\pm$          96& -1.81$\pm$  0.11&  0.73$\pm$  0.31&-  &0.80\\
J132517.5-472427&        4265$\pm$         101& -2.00$\pm$  0.10&  0.51$\pm$  0.35& -0.04  & 0.48\\ 
J132521.3-473654&        4000$\pm$          75& -1.75$\pm$  0.09&  0.50$\pm$  0.20&-  &0.80\\
J132545.2-473238&        4073$\pm$          96& -1.45$\pm$  0.11&  0.61$\pm$  0.31&+0.05  &0.60\\
J132551.2-472702&        4497$\pm$         101& -2.00$\pm$  0.10&  0.50$\pm$  0.35&-  &0.70\\
J132552.0-473016&        4250$\pm$          96& -1.75$\pm$  0.11&  0.50$\pm$  0.31& +0.13 & 0.80\\ 
J132558.7-473610&        4256$\pm$          86& -0.64$\pm$  0.10&  1.30$\pm$  0.20&-  &0.52\\
J132601.7-474034&        4497$\pm$          96& -1.85$\pm$  0.11&  0.30$\pm$  0.31&-0.02 &0.79\\
J132609.1-472720&        4493$\pm$         101& -2.00$\pm$  0.10&  0.50$\pm$  0.35&-  &0.90\\
J132614.6-472123&        4279$\pm$          96& -1.75$\pm$  0.11&  0.61$\pm$  0.31&-0.08  &0.67\\
J132623.7-474243&        4251$\pm$          96& -1.20$\pm$  0.11&  0.60$\pm$  0.31&+0.32  &0.82\\
J132629.6-473701&        4298$\pm$          96& -1.44$\pm$  0.11&  0.01$\pm$  0.31&-0.23  &0.64\\
J132639.0-474359&        4499$\pm$          86& -0.75$\pm$  0.10&  2.50$\pm$  0.20&-  &0.59\\
J132639.3-472035&        4250$\pm$          75& -1.75$\pm$  0.09&  0.50$\pm$  0.20&+0.11  &0.65\\
J132646.2-471415&        4466$\pm$          75& -1.78$\pm$  0.09&  0.93$\pm$  0.20&+0.11  &0.71\\
J132654.3-474605&        4575$\pm$          76& -2.15$\pm$  0.10&  0.73$\pm$  0.27&+0.13  &0.72\\
J132704.4-443003&        4881$\pm$          96& -1.51$\pm$  0.11&  1.96$\pm$  0.31&-0.06  &0.76\\
J132709.6-472052&        4258$\pm$          76& -2.00$\pm$  0.10&  0.51$\pm$  0.27&-  &0.76\\
J132710.5-473701&        4000$\pm$          96& -1.29$\pm$  0.11&  0.42$\pm$  0.31& +0.14  &0.81\\
J132726.0-473060&        4500$\pm$          96& -1.75$\pm$  0.11&  1.00$\pm$  0.31&-  &0.79\\
J132753.7-472442&        4577$\pm$          86& -0.76$\pm$  0.10&  0.44$\pm$  0.20&-  &0.74\\
J132754.7-471932&        3891$\pm$          69& -0.42$\pm$  0.09&  4.51$\pm$  0.15&-  &0.52\\
J132757.3-473638&        4669$\pm$         101& -2.09$\pm$  0.10&  1.23$\pm$  0.35&-  &0.84\\
J132800.8-473247&        4587$\pm$          76& -2.16$\pm$  0.10&  0.81$\pm$  0.27&-  &0.71\\
J132804.8-474504&        4251$\pm$          96& -1.25$\pm$  0.11&  2.00$\pm$  0.31&-  &0.79\\
J132813.6-472424&        4000$\pm$          96& -1.50$\pm$  0.11&  0.00$\pm$  0.31&-  &0.83\\
J132815.0-473739&        4249$\pm$          96& -1.25$\pm$  0.11&  1.00$\pm$  0.31&-  &0.74\\
J132816.9-472956&        4311$\pm$         101& -2.04$\pm$  0.10&  0.79$\pm$  0.35&-0.07  &0.82\\
J132833.8-473206&        4250$\pm$          75& -1.75$\pm$  0.09&  0.50$\pm$  0.20&-0.10  &0.71\\	
J132839.9-472633&        4500$\pm$         112&  0.00$\pm$  0.11&  2.00$\pm$  0.24&-  &0.66\\
J132918.9-471924&        4250$\pm$          75& -1.75$\pm$  0.09&  0.50$\pm$  0.20&-  &0.70\\
J132936.8-500005&        4500$\pm$          96& -1.25$\pm$  0.11&  2.00$\pm$  0.31&-  &0.61\\
J133106.0-483312&        3801$\pm$          96& -1.00$\pm$  0.11&  1.00$\pm$  0.31&-  &0.68\\
J133328.7-441903&        4249$\pm$          75& -1.49$\pm$  0.09&  0.51$\pm$  0.20&-  &0.85\\
J133939.4-490014&        4689$\pm$          62& -0.99$\pm$  0.08&  1.88$\pm$  0.14&-0.01  &0.52\\
J134508.6-492832&        3800$\pm$          96& -1.25$\pm$  0.11&  0.50$\pm$  0.31& +0.03  & 0.58\\
\hline
\label{tab:NGC5139_stelpars}
\end{tabular}
\end{minipage} 
\end{table*}

 \begin{figure}
  \centering  
   \includegraphics[width=1.00\linewidth]{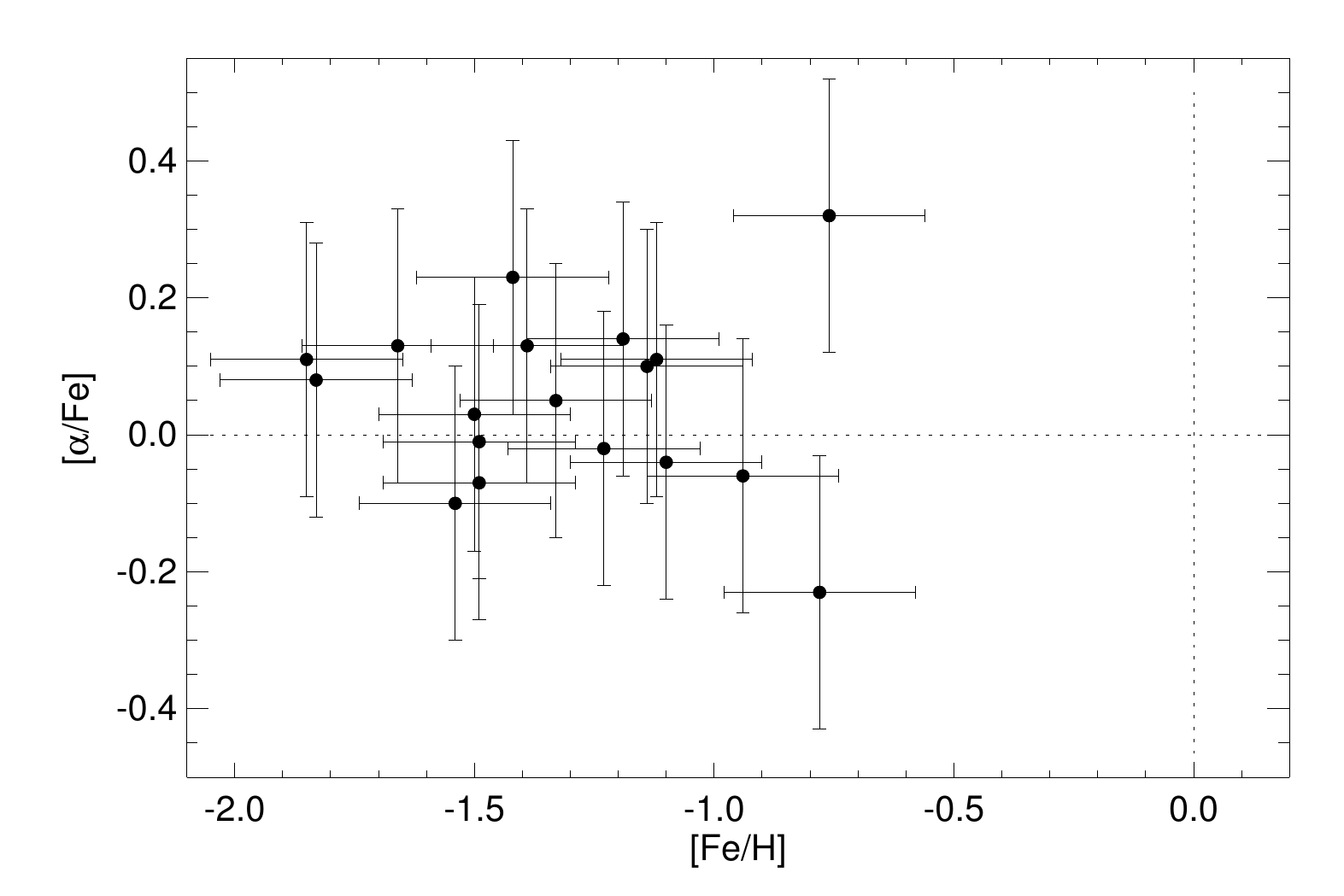}
  \caption{[Fe/H] - [$\alpha$/Fe] diagram for stars selected as likely members of $\omega$ Cen, plotting the abundances from the RAVE chemical catalogue \citep{2011AJ....142..193B}. We found a large spread in [$\alpha$/Fe] ranging from -0.23 to +0.32 dex, with a measurement error of $\sigma_{[\alpha/Fe]} \sim$ 0.2 dex.}
  \label{fig:NGC5139_alpha}
\end{figure} 

\citet{2010ApJ...722.1373J}, using 855 red giant stars, found that the $\alpha$ elements in $\omega$ Cen are generally enhanced by $\sim$ +0.3 dex and exhibit a metallicity-dependent morphology. 
We find a large spread in [$\alpha$/Fe] for a given [Fe/H] (see Fig.\ref{fig:NGC5139_alpha} and Table \ref{tab:NGC5139_stelpars}) using the values from the RAVE chemical abundance catalogue \citep{2011AJ....142..193B}. The measured [$\alpha$/Fe] ranges from -0.23 to +0.32 dex with a mean error of $\sim$0.2 dex \citep{2011AJ....142..193B}. \citet{2002ApJ...568L.101P} and \citet{2007ApJ...663..296V} also found a large spread in $\alpha$-elements for a given [Fe/H]. 

 \begin{figure}
  \centering  
   \includegraphics[width=1.50\linewidth]{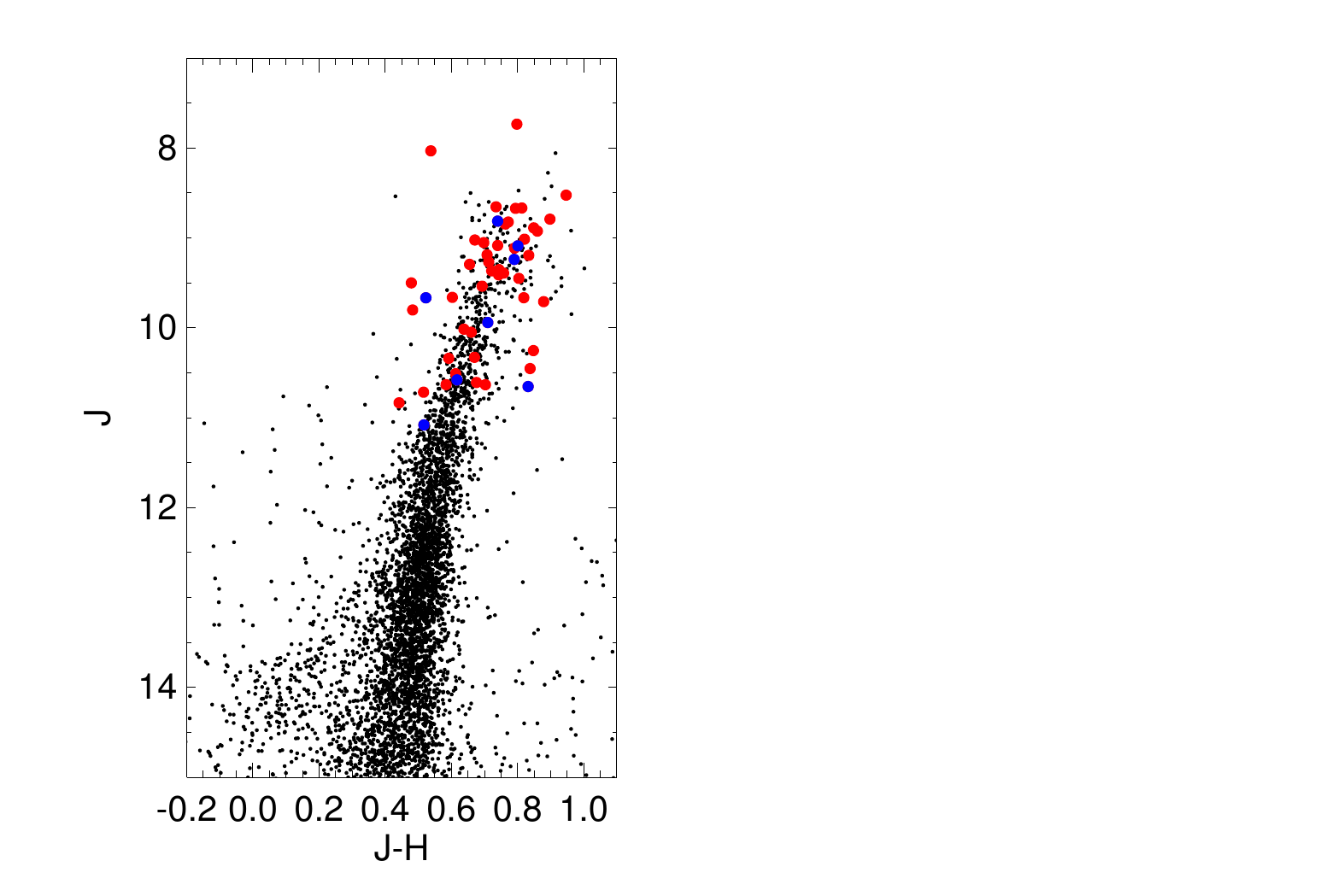}
  \caption{2MASS J, J-H Colour-Magnitude Diagram (CMD) centred on $\omega$ Cen, with a radius of 0.1 degrees. The big dots are the stars selected for being likely members of the cluster based on RV and location within the tidal radius; stars with large proper motions with respect to the cluster are shown in blue. Most of the stars selected using RV and location clearly lie on or near the prominent RGB of $\omega$ Cen.}
  \label{fig:NGC5102_CMD}
\end{figure} 

\begin{figure*}
  \centering  
   \includegraphics[width=0.80\linewidth]{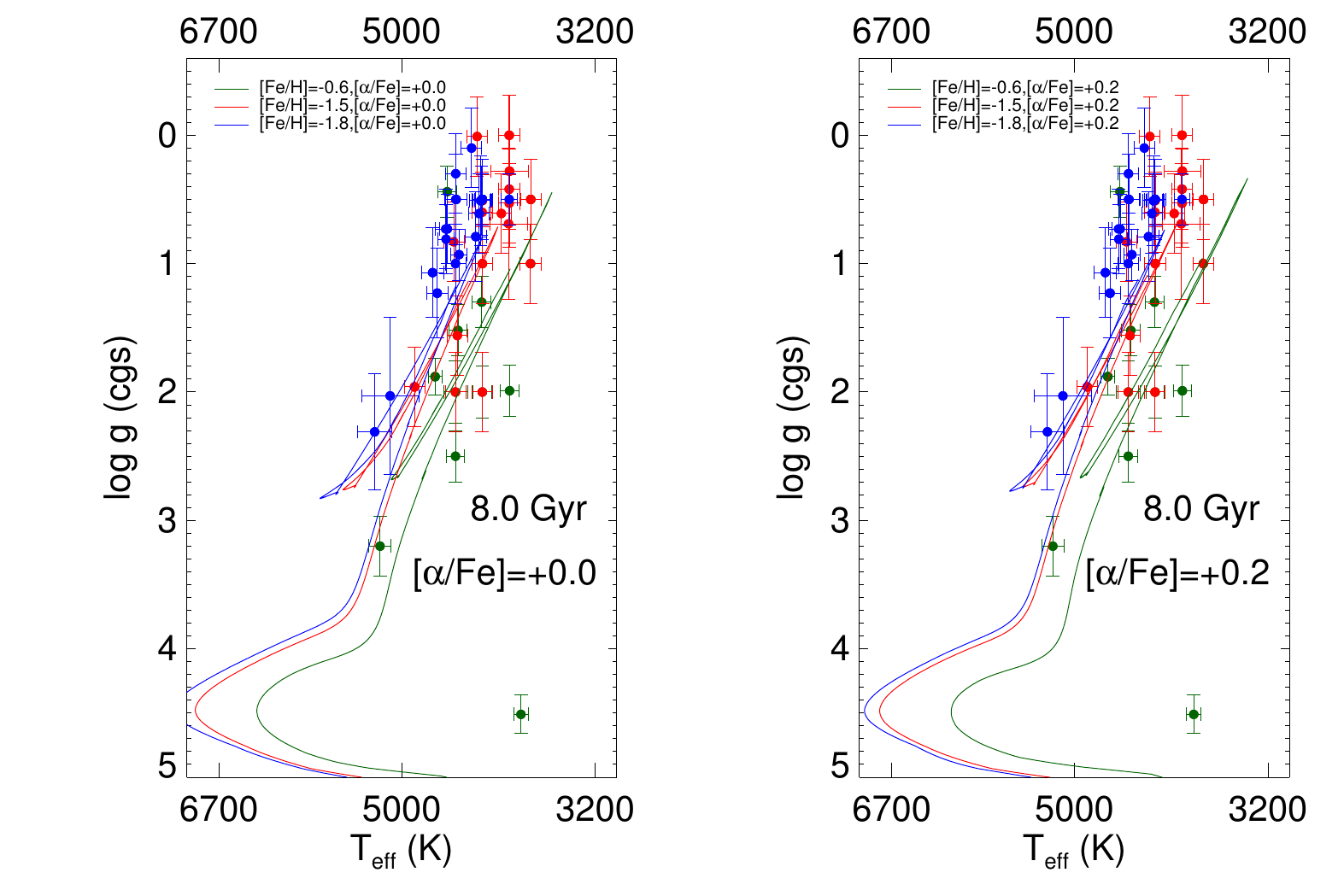}
   \includegraphics[width=0.80\linewidth]{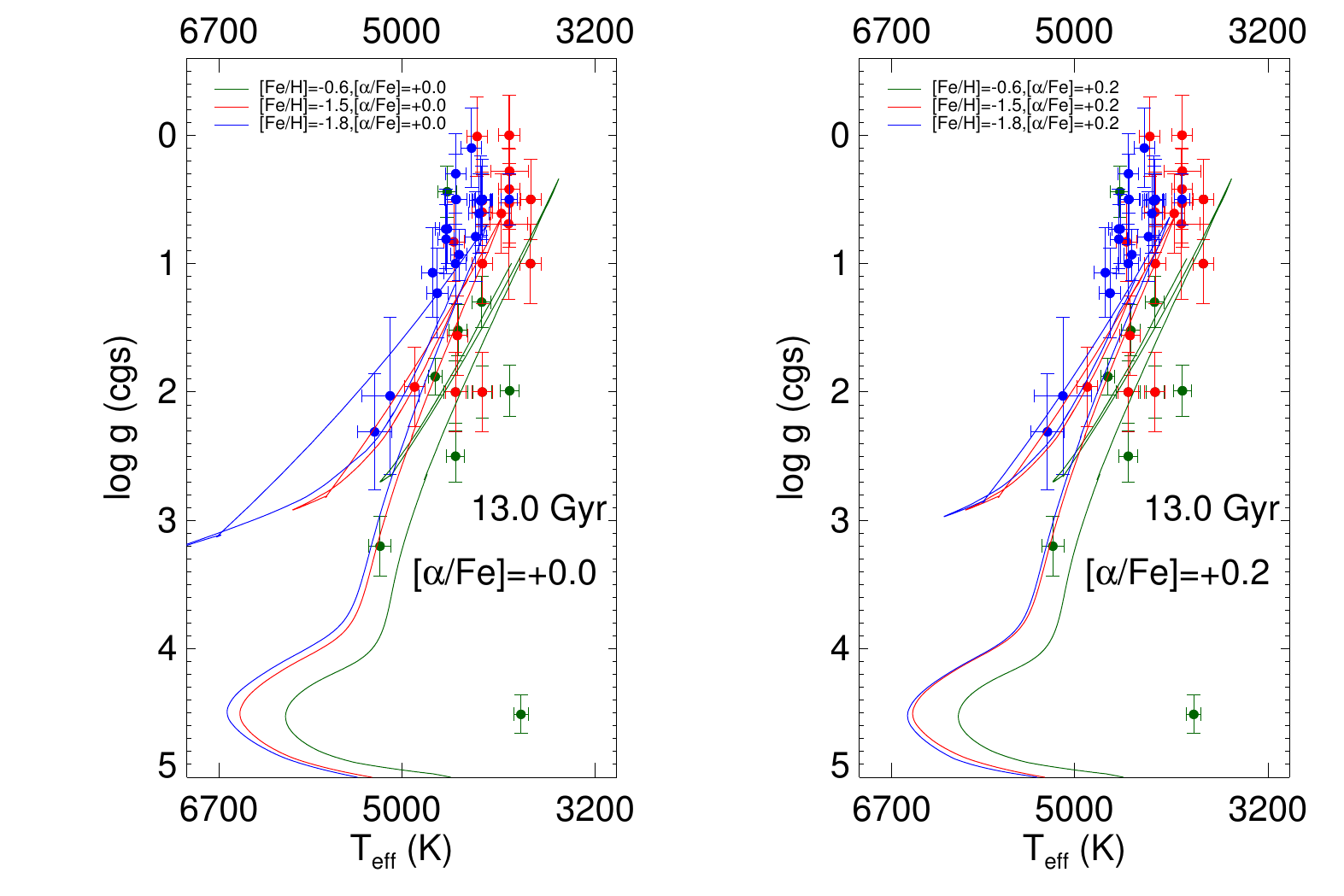}
  \caption{Temperature-surface gravity diagram for the stars selected as likely members of NGC 5139 ($\omega$ Cen). In the top panels we plot BASTI isochrones covering the range in [Fe/H] from -1.8 to -0.6 for [$\alpha$/Fe] = +0.0 ({\emph left}) and +0.2 ({\emph right}) at an age of 8 Gyr.  In the bottom panel we use the same set of isochrones, with the same range of metallicities and alpha enhancements, but at an age of 13 Gyr. The green dots represent stars with metallicities between 0.0 $<$ [Fe/H] $<$ -1.0, the red dots are stars with -1.0 $<$ [Fe/H] $<$ -1.6, and the blue dots are those with -1.6 $<$ [Fe/H] $<$ -2.2\,.}
  \label{fig:teff_logg_wCen}
\end{figure*} 

An age spread from 8 to 13 Gyr between different members of the cluster was found by \citet{2007ApJ...663..296V}. In Fig.~\ref{fig:teff_logg_wCen} we show a T$_{\rm eff}$ - $\log$ g plot of the high probability $\omega$ Cen members found in the RAVE survey, with the data colour-coded by RAVE [Fe/H]: green dots represent stars between 0.0 $<$ [Fe/H] $<$ -1.0 (note that there is only one star at approximately solar metallicity); the red dots are stars with -1.0 $<$ [Fe/H] $<$ -1.6;  and the blue dots are stars with -1.6 $<$ [Fe/H] $<$ -2.2. We overplotted BASTI isochrones with [Fe/H] = [-0.6, -1.5, -1.8], and [$\alpha$/Fe] = [+0.0, +0.2],  for ages of 8.0 Gyr (top panels) and 13 Gyr (bottom panels). (Fig.\ref{fig:NGC5102_meta} appears to show three [Fe/H] peaks in our data, at -0.6, -1.3 and -1.8 dex, hence our selection of these isochrones.) There is a reasonable match (within the errors) between the $\log$ g, T$_{\rm eff}$ and [Fe/H] values derived from the RAVE spectra and the selected isochrones, at least for the more metal-rich stars. However, an isochrone younger than 8.0 Gyr would be a better fit, especially for the most metal-poor stars (red and blue dots). \citet{2007ApJ...663..296V} found that the youngest stars in the cluster are around 8.0 Gyr. If this age limit is correct, and the most metal-poor stars in $\omega$ Cen are not somehow younger than their more metal-rich counterparts, the isochrones would suggest that the spectroscopic gravities for the metal-poor stars are underestimated. Note also that for the more metal-rich the underestimation in the spectroscopic gravities is less evident. 

\subsubsection{RAVE stellar parameters and the \citet{2010ApJ...722.1373J} study}

We identified 21 stars in common between the $\omega$ Cen candidates found in RAVE and the high-resolution, high S/N spectra of 855 $\omega$ Cen RGB members obtained and analysed for elemental abundances by \citet{2010ApJ...722.1373J}.  The sample includes nearly all RGB stars brighter than V = 13.5 and spans $\omega$ Cen's full metallicity range.

\begin{figure*}
  \centering  
   \includegraphics[width=0.45\linewidth]{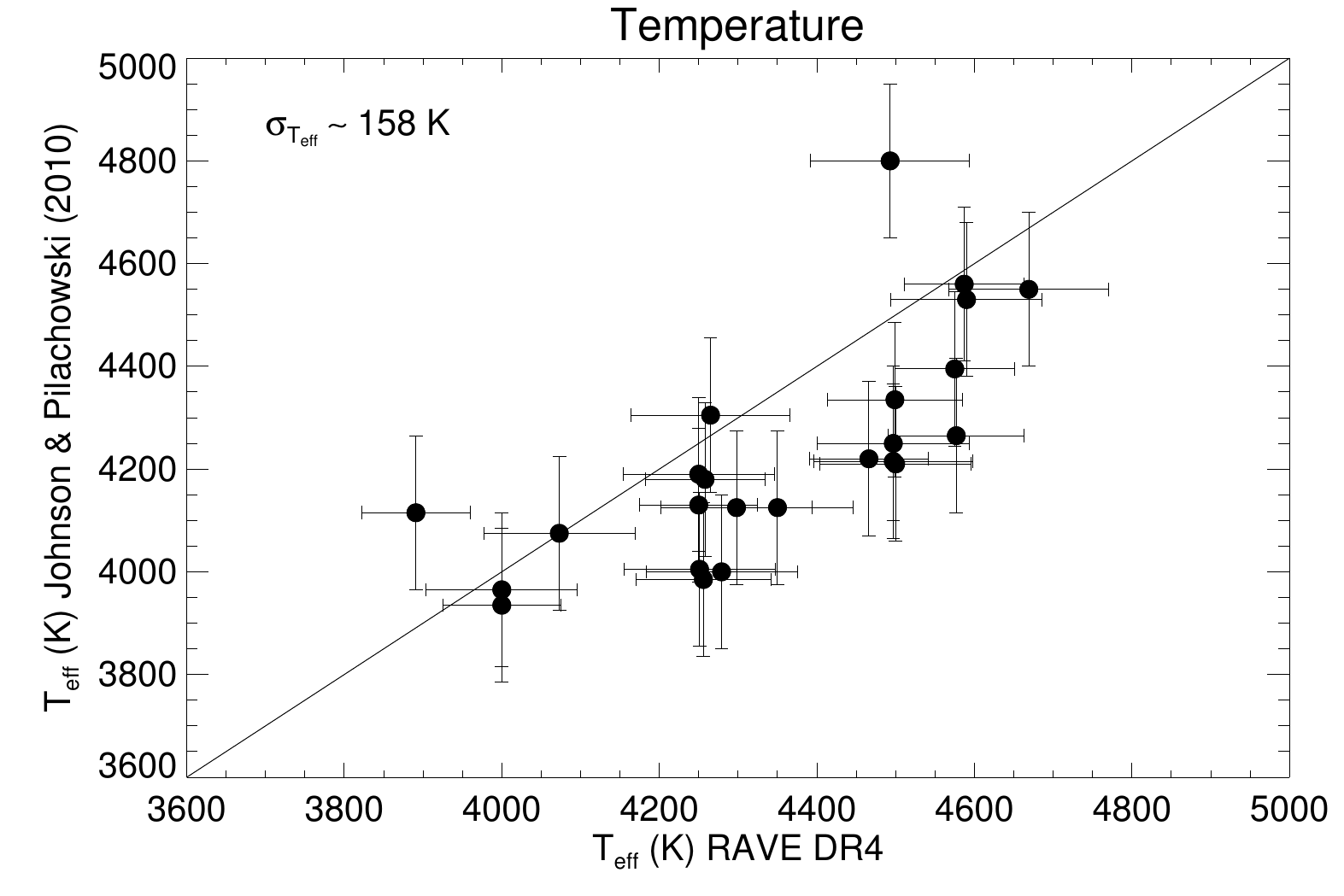}
   \includegraphics[width=0.45\linewidth]{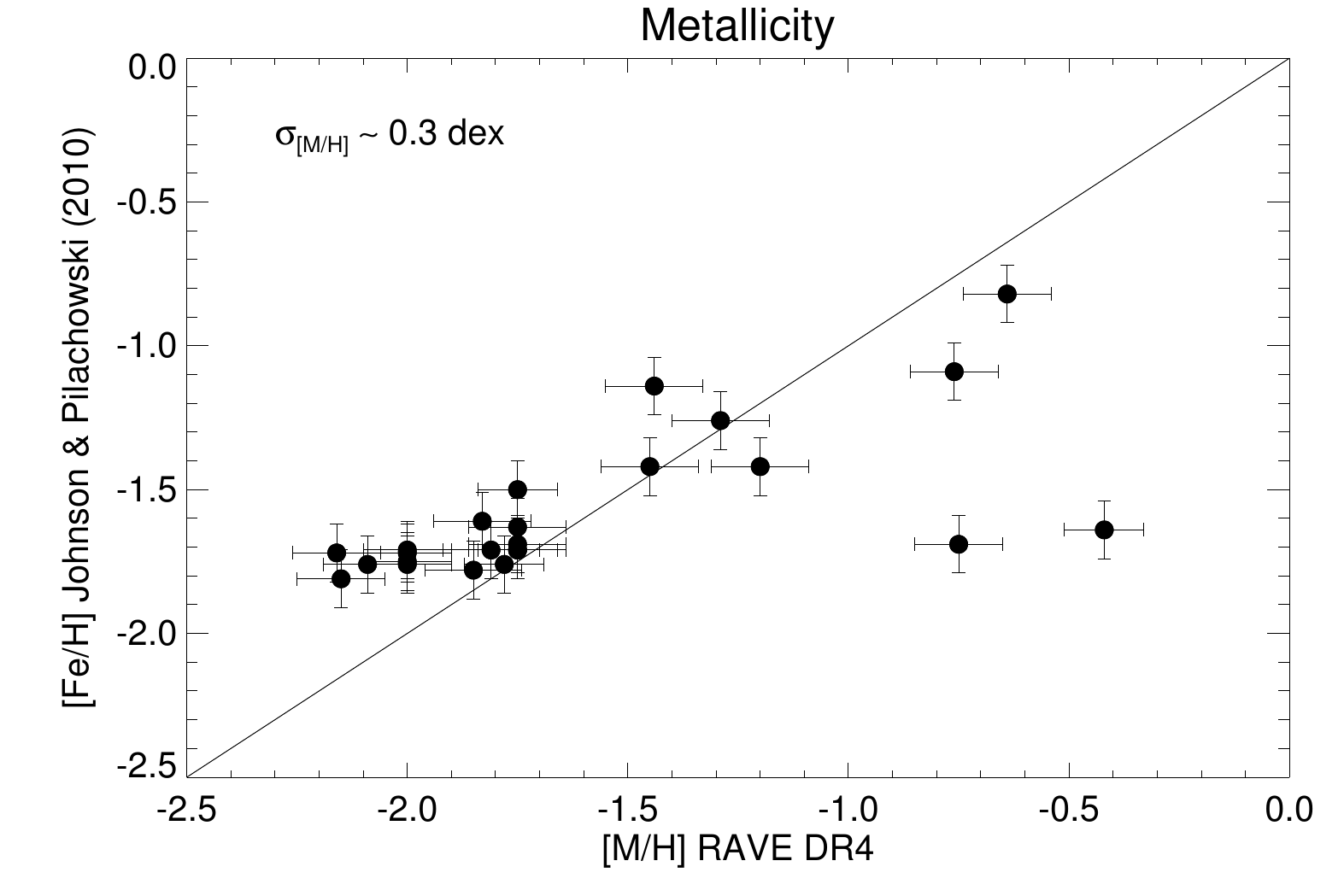}
   \includegraphics[width=0.45\linewidth]{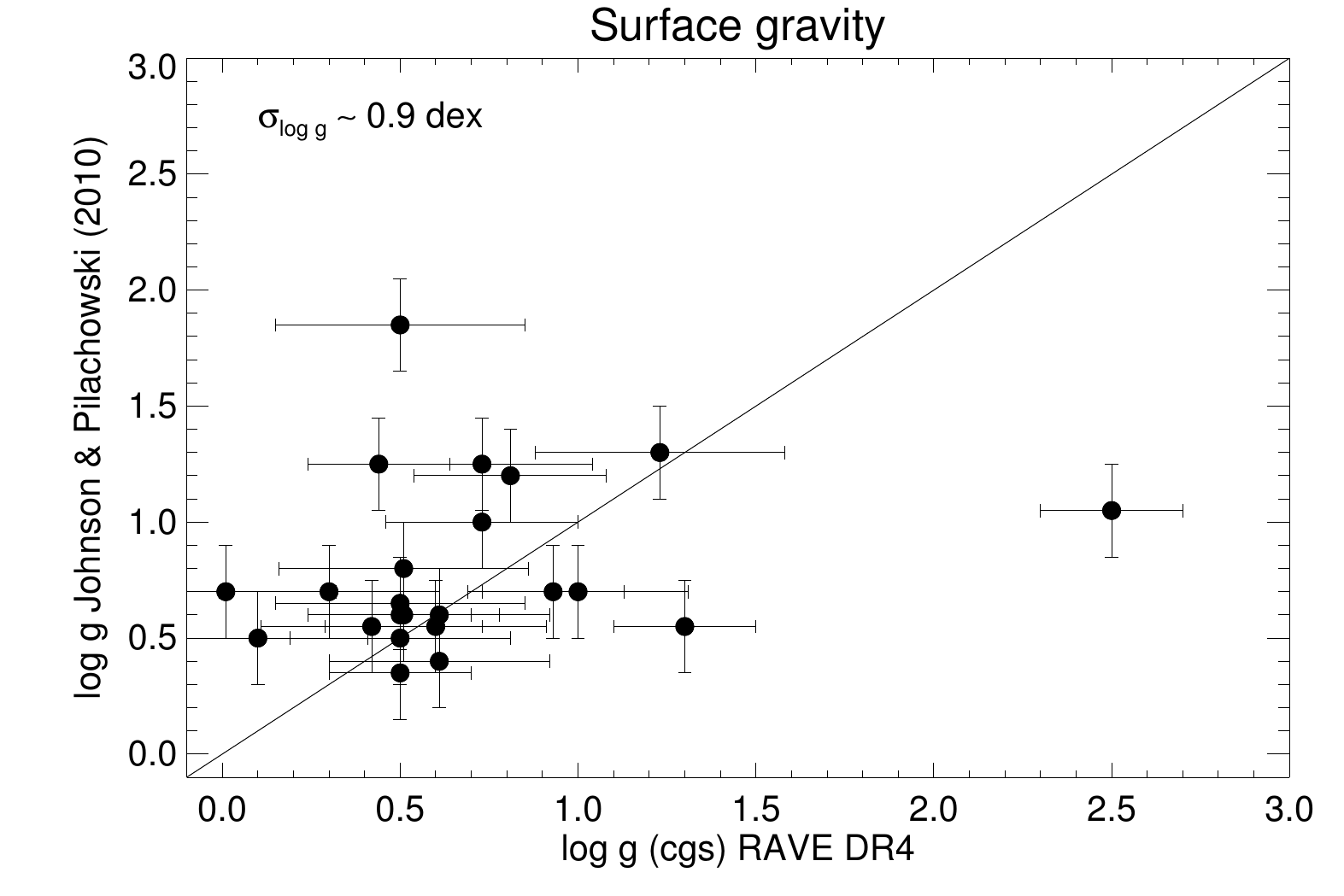}
   \includegraphics[width=0.45\linewidth]{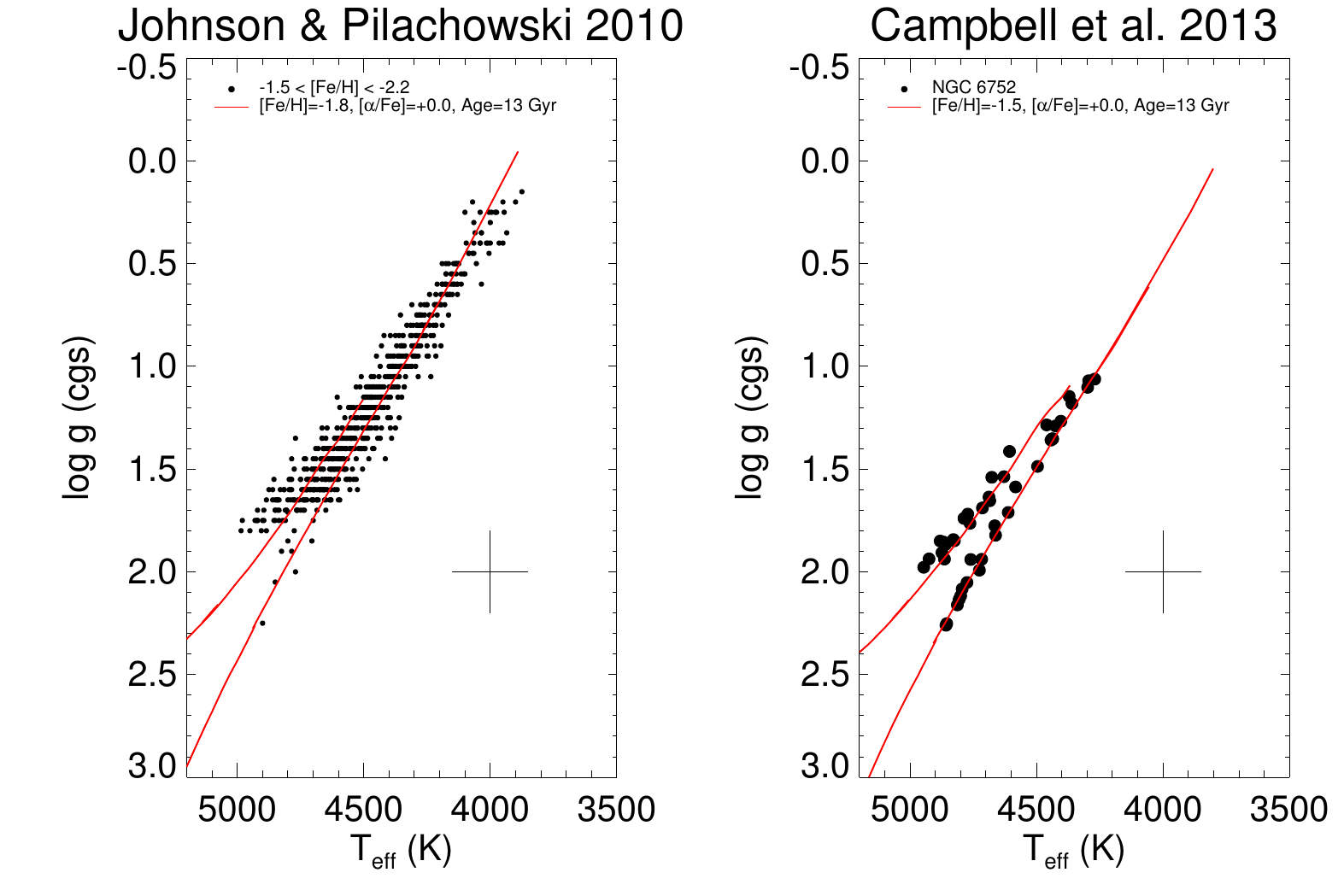}
    \caption{1:1 comparison for 21 stars in common between the RAVE DR4 catalogue and the data-set from \citet{2010ApJ...722.1373J}. We compared T$_{\rm eff}$, metallicity and surface gravity, respectively. The lower right panel show data from \citet{2010ApJ...722.1373J} for $\omega$ Cen and \citet{2013Natur.498..198C} for the cluster NGC 6752. BASTI isochrones are also showed (Fe/H] = -1.8, -15 dex respectively). We found a good agreement between these data-sets and the isochrones.}
  \label{fig:stellar_param_NGC5139}
\end{figure*} 

Effective temperatures (T$_{\rm eff}$) were determined via the empirical (V -- K) color--temperature relation from \citet{1999A&AS..140..261A} using the recommended value of E(B -- V) = 0.12 \citep{1996AJ....112.1487H}. Surface gravity estimates were obtained using the photometric temperatures and absolute bolometric magnitudes (M$_{\rm bol}$), assuming stellar masses of 0.8 M$_{\odot}$. Chemical abundances were determined through standard equivalent width (EW) analyses using the LTE line analysis code MOOG. We compared the temperature, metallicity and surface gravity derived from the RAVE spectra and those derived in \citet{2010ApJ...722.1373J} for the 21 stars in common (see Fig.~\ref{fig:stellar_param_NGC5139}). While there is a general correlation in T$_{\rm eff}$, we found a systematic offset, such that RAVE temperatures are slightly hotter with respect to the photometric ones used in \citet{2010ApJ...722.1373J}. For metallicities we also found a rough correlation between both methods, although there are two clearly discrepant stars, with RAVE metallicities $\sim 1$ dex higher than the high-resolution abundances. Finally, surface gravities in the small range where these giants lie also appear to be offset, with RAVE estimates mostly in the range log(g) $\sim$0.0 -- 1.0 and \citet{2010ApJ...722.1373J} values spanning 0.5 -- 1.5. The standard deviations of the differences in T$_{\rm eff}$, [M/H] and log g are 158 K, 0.3 dex and 0.9 dex respectively. Isochrones are affected by many uncertainties in their underlying physics associated with e.g., treatment of convection, surface boundary conditions, etc. We found discrepancies between the stellar parameters derived from the RAVE spectra and the BASTI isochrones used in this work (see Fig.~\ref{fig:NGC3201_CMD} and Fig.~\ref{fig:teff_logg_wCen}). The lower right panel of Fig.~\ref{fig:stellar_param_NGC5139} shows data from \citet{2010ApJ...722.1373J} for $\omega$ Cen and \citet{2013Natur.498..198C} for NGC 6752 together with BASTI isochrones with [Fe/H] = -1.8 dex and -1.5 dex, respectively. We found good agreement between the isochrones and the data from these two independent studies, in contrast to the offset between RAVE pipelines results and the BASTI isochrones noted above.

\subsubsection{Stellar distances}



\begin{figure*}
  \centering  
   \includegraphics[width=0.80\linewidth]{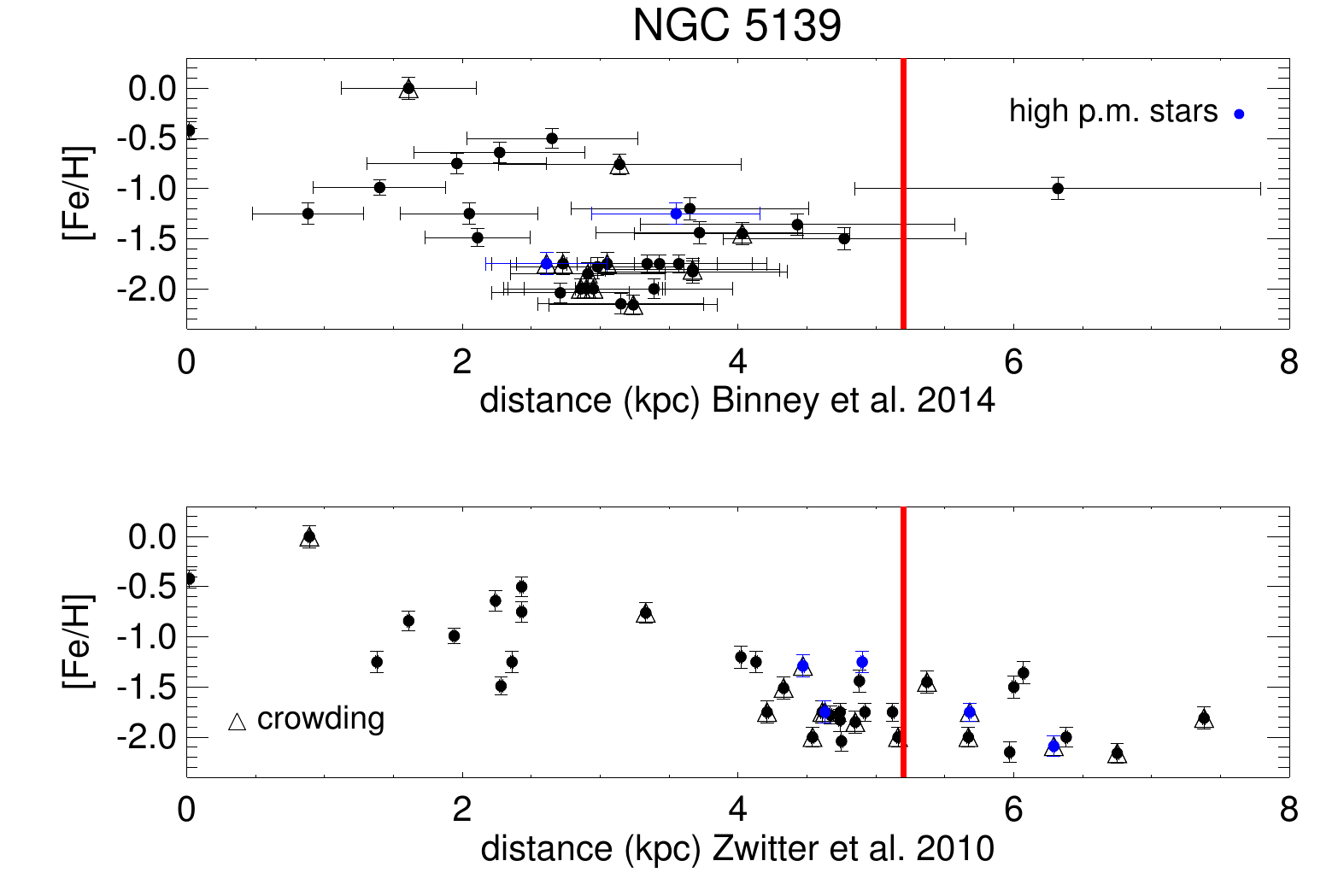}
    \caption{Distance vs. [Fe/H] for the members of NGC 5139 ($\omega$ Cen). In the top panel the distances derived in \citet{2013MNRAS.tmp.2584B} are shown; in the bottom panel the distances are from \citet{2010A&A...522A..54Z}. Blue dots are stars with a large proper motion relative to $\omega$ Cen; triangles indicate stars affected by crowding. The vertical red line indicates the distance of the cluster from \citet{1996AJ....112.1487H} (2010 edition).}
  \label{fig:NGC5139_dist}
\end{figure*}

Here we adopt the true distance modulus from \citet{1996AJ....112.1487H} (2010 edition), (m-M)$_{0}$ = 13.58, i.e., 5.2 kpc. Fig.\ref{fig:NGC5139_dist} shows the stellar distances for NGC 5139 members together with their measured [Fe/H]; the top panel shows the distances from \citet{2013MNRAS.tmp.2584B} and the bottom panel the distances calculated by \citet{2010A&A...522A..54Z}. The blue dots in the figure are stars with high measured proper motions with respect to the nominal value for the cluster as described above. \citet{2013MNRAS.tmp.2584B} found a wide range of distances for these stars, from 1 to 7 kpc. Most of the stars are between 2 and 4 kpc with a peak at $\sim$ 3 kpc. The most metal-poor stars show a small dispersion in distances. \citet{2010A&A...522A..54Z} also found stars ranging from 1 to 7 kpc, but they show a fairly clear trend between [Fe/H] and distances, with metal-rich stars estimated to be closer than metal-poor ones. It is worth noting that, despite the wide spread in measured distances, \citet{2010A&A...522A..54Z} found a group of stars with a mean distance around 5 kpc and [Fe/H] between -1.0 and -2.2 dex. This value is in good agreement with the distances for $\omega$ Cen found in the literature. When applied to these stars, the aforementioned new distance calibration using isochrones extending to [Fe/H] $< -0.9$ gives a mean distance of 4.2 $\pm$ 1.2 kpc.

There is also evidence of field stars that may be associated with $\omega$ Cen in the nearby Galactic disc \citep[e.g.,][]{2010AJ....139..636W}. If the distances from \citet{2013MNRAS.tmp.2584B} are correct, that might suggest that we found a former part of the cluster that is closer to us than the cluster itself. However, if these stars are in fact part of the main cluster the result would suggest that the distances from \citet{2013MNRAS.tmp.2584B} are underestimated by $\sim$ 40$\%$. Note that using the distances from \citet{2010A&A...522A..54Z}, we find that most of the stars are between 4 and 6 kpc, in good agreement with distances determined for this cluster from the literature.  



\subsection{The case of NGC 362} 

NGC 362 has a metallicity [Fe/H] = -1.26, according to the 2010 update of the \citet{1996AJ....112.1487H} catalogue. The orbit has a high eccentricity and a low inclination, and is confined close to the Galactic plane \citep{1999AJ....117.1792D}. This cluster shows a split in the red giant branch. Recently, \citet{2013A&A...557A.138C} analysed FLAMES GIRAFFE+UVES spectra for 92 stars in the cluster and found that stars seem to be clustered into two discrete groups along the Na-O anti-correlation. \citet{2013A&A...557A.138C} did not find a significant spread in [Fe/H], with the star-to-star variation being $\sim$ 0.1 dex and the mean [Fe/H] $\sim$ -1.2. \citet{2008A&A...486..437K} found a clear bi-modality in CN in NGC 362 and \citet{2010MNRAS.406.2504W} found homogeneity in s- and r-process abundances. \citet{2009ApJ...694.1498M} used deep and homogeneous photometry from HST to derive an age of 10.3 Gyr for NGC 362. We selected high probability members of NGC 362 from the RAVE catalogue by taking into account the stars' RVs, proper motions and location within the tidal radius of the cluster. In the next section we explore the main stellar parameters for these objects. 

\subsubsection{Abundances, temperatures and gravities}

\begin{figure*}
  \centering  
   \includegraphics[width=0.80\linewidth]{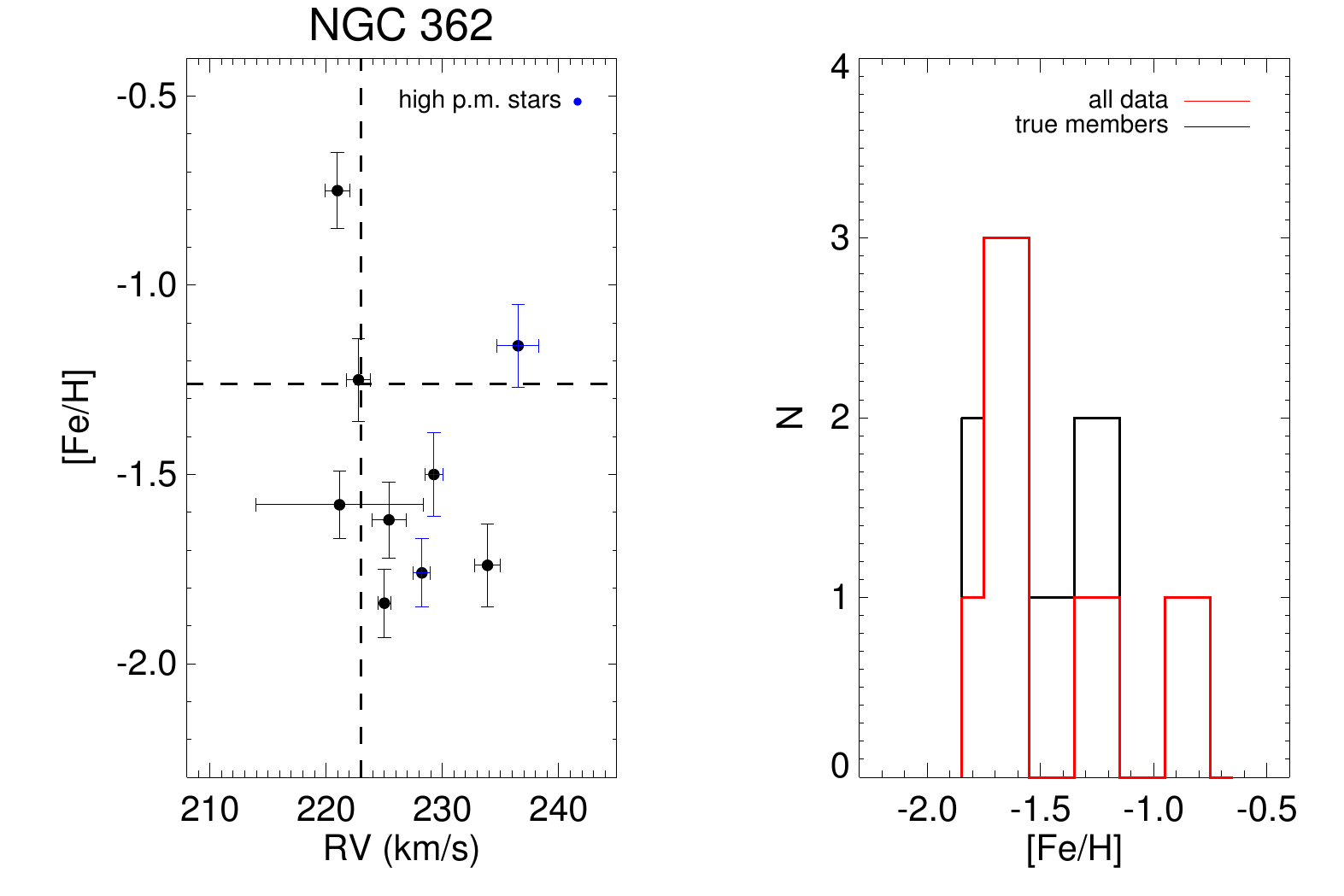}
    \caption{\emph{Left panel:} Radial velocity-abundance plot for RV candidate members which fall within the tidal radius of NGC 362. The dashed lines indicate the nominal RV and [Fe/H] for the cluster reported in the literature. The blue dots are stars with large proper motions with respect to the measured proper motion for the cluster. \emph{Right panel:} The metallicity distribution function for all RV candidate members which fall within the tidal radius (in red) and for those which also have proper motion measurements consistent with that of the cluster (black line).}
  \label{fig:RV_meta_NGC362}
\end{figure*} 

Most of the stars identified as the highest likelihood members for this cluster have measured metallicities [Fe/H] $\sim$ -1.7\,. We also find stars with [Fe/H] = -1.2 and -0.7  at the nominal RV for NGC 362 (see Fig.~\ref{fig:RV_meta_NGC362}). Blue dots in Fig.~\ref{fig:RV_meta_NGC362} are stars with RVs similar to the cluster but they have large proper motions with respect to NGC 362. From the [Fe/H] distribution function in the right panel of Fig.~\ref{fig:RV_meta_NGC362} most of the stars have [Fe/H] between -1.5 and -1.8. As noted above, high-resolution spectroscopy of members of this cluster has not yielded a significant spread in [Fe/H],  although \citet{1981BAAS...13S.545P} derived a mean [Fe/H] = -0.9 from several giants while \citet{2000AJ....119..840S} obtained [Fe/H] = -1.33 from 12 giants in this cluster. The large range in metallicity for candidate NGC 362 members from RAVE -- in particular the apparent clump at [Fe/H] $\sim$ -1.7 -- is thus somewhat puzzling. Barring a significant spread in metallicity not detected in previous work, contamination by field stars and/or bright stars from the SMC would seem to be the most likely explanation for the observed abundances in our sample.  

\begin{figure*}
  \centering  
   \includegraphics[width=0.80\linewidth]{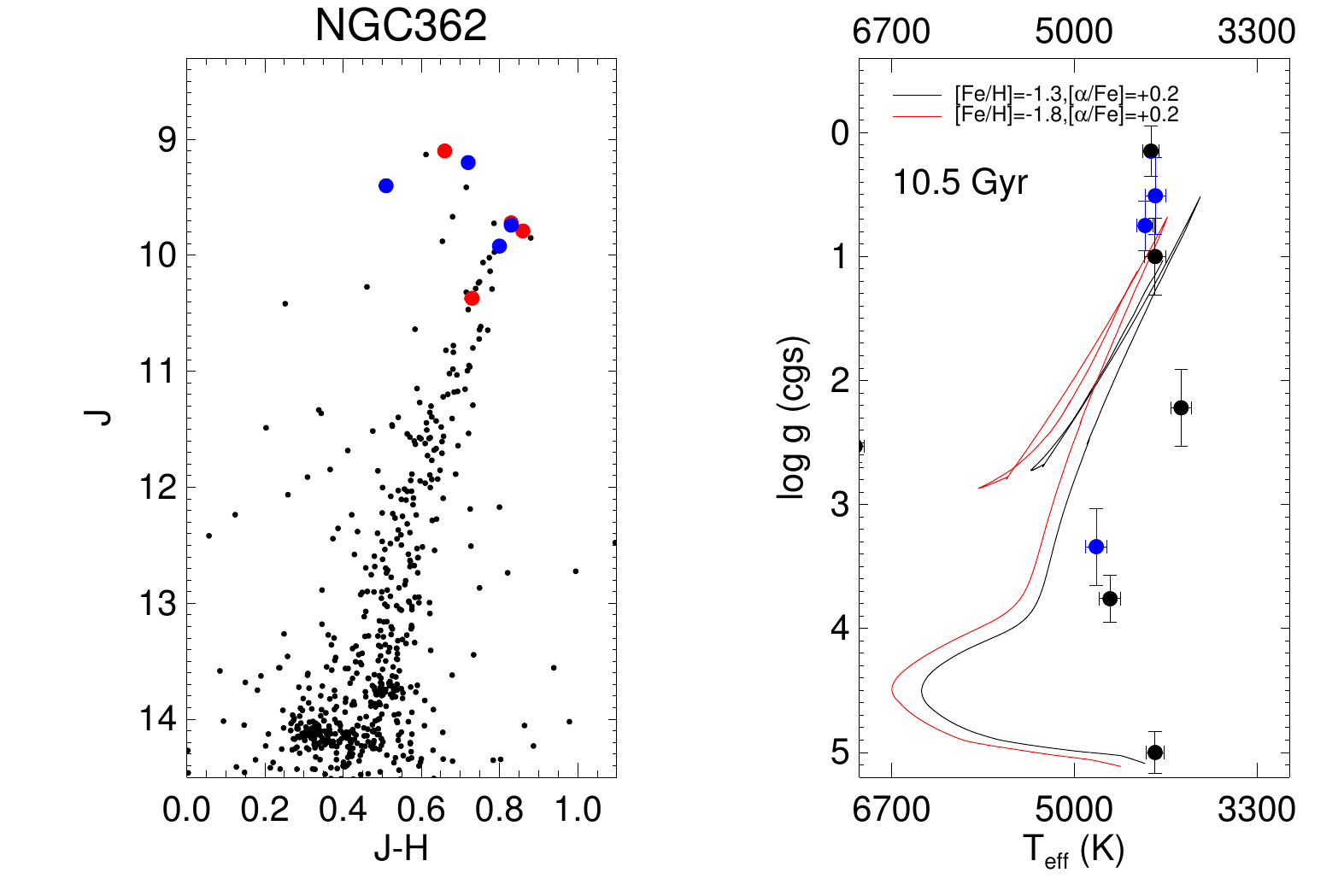}
    \caption{\emph{Left panel:} 2MASS J, J-H Colour-Magnitude Diagram (CMD) centred on NGC 362 with a radius of 0.1 degrees. The large dots are the stars selected as being probable members of the cluster based on RV and location within the tidal radius; stars with large proper motions with respect to the cluster are in blue. Note that five stars are found on the RGB while 3 stars are clearly outliers, suggesting they are not members of the cluster. \emph{Right panel:} Temperature-surface gravity diagram for the stars selected as probable members of NGC 362, overplotted with BASTI isochrones covering the range in [Fe/H] from -1.3 to -1.8 and [$\alpha$/Fe] = +0.2 dex at an age of 10.5 Gyr. Four stars lie relatively close to the selected isochrones. Stars with large proper motions relative to the cluster are shown as blue dots.}
  \label{fig:teff_logg_NGC362}
\end{figure*}

In Fig.~\ref{fig:teff_logg_NGC362} we show a CMD and temperature-surface gravity diagram of the candidate members of NGC 362, overplotted with BASTI isochrones for [Fe/H] = -1.3 dex \citep{2009A&A...508..695C} and an age of 10.5 Gyr (e.g. \citealt{2009ApJ...694.1498M}). In the left panel, five stars lie in the region of the RGB; of these five, four stars also fall near the giant branch of the isochrones in the right panel. However, as with NGC 3201 and $\omega$ Cen, the position of these stars relative to the isochrones would suggest that surface gravities in the RAVE database might be underestimated for the given metallicity and age of the cluster.

\begin{table*}
 \centering
 \begin{minipage}{370mm}
  \caption{NGC 362 candidates selected from the RAVE data (RV, proper motions, tidal radius) and their parameters}
  \begin{tabular}{@{}rrrrrrrrrrrrr@{}}
  \hline
  \hline
ID & T$_{eff}$ (K) & [Fe/H] & $\log$ g (cgs) & [$\alpha$/Fe] & J-H    \\
     \\
 \hline
J004905.3-733108&        4000$\pm$          96& -1.25$\pm$  0.11&  0.50$\pm$  0.31& -&0.55\\
J004217.1-740615&        4526$\pm$         101& -2.68$\pm$  0.10&  0.12$\pm$  0.35&  -&0.38\\
J005038.4-732818&        7039$\pm$          83& -1.58$\pm$  0.09&  2.53$\pm$  0.13&   - & -\\
J010313.6-705037&        4250$\pm$          84& -0.75$\pm$  0.10&  5.00$\pm$  0.17& - &0.50\\
J010314.7-705115&        4669$\pm$          98& -1.62$\pm$  0.10&  3.76$\pm$  0.19& -&0.32\\
J010314.7-705059&        4288$\pm$          75& -1.84$\pm$  0.09&  0.15$\pm$  0.20& -&0.53\\
J010315.1-705032&        4008$\pm$          96& -1.74$\pm$  0.11&  2.22$\pm$  0.31& -&0.84\\
J010319.0-705051&        4250$\pm$          96& -1.25$\pm$  0.11&  1.00$\pm$  0.31& -&0.61\\
J010335.7-705052&        4000$\pm$          96& -1.50$\pm$  0.11&  0.00$\pm$  0.31&  -&0.29\\
J011655.9-690607&        4033$\pm$         173& -1.21$\pm$  0.21&  0.59$\pm$  0.59&  -&0.57\\
\hline
\label{tab:NGC362_stelpars}
\end{tabular}
\end{minipage}
\end{table*}

Table \ref{tab:NGC362_stelpars} lists the stellar parameters derived in the RAVE survey database for the NGC 362 candidates. Unfortunately, there are no  [$\alpha$/Fe] measurements in the RAVE chemical abundance catalogue for these objects.  

We adopt the distance modulus given in \citet{1996AJ....112.1487H} (2010 edition), (m-M)$_{V}$ = 14.83 using a E(B-V) = 0.05 for this globular cluster. Fig.~\ref{fig:teff_logg_NGC362} shows a few stars clearly outside of the isochrones for the metallicity and age reported for this cluster, suggesting that these objects are halo field stars (or members of the SMC) with a RV similar to NGC 362. High resolution follow-up spectroscopic observations of these targets would help to settle this question.  


\begin{figure*}
  \centering  
   \includegraphics[width=0.80\linewidth]{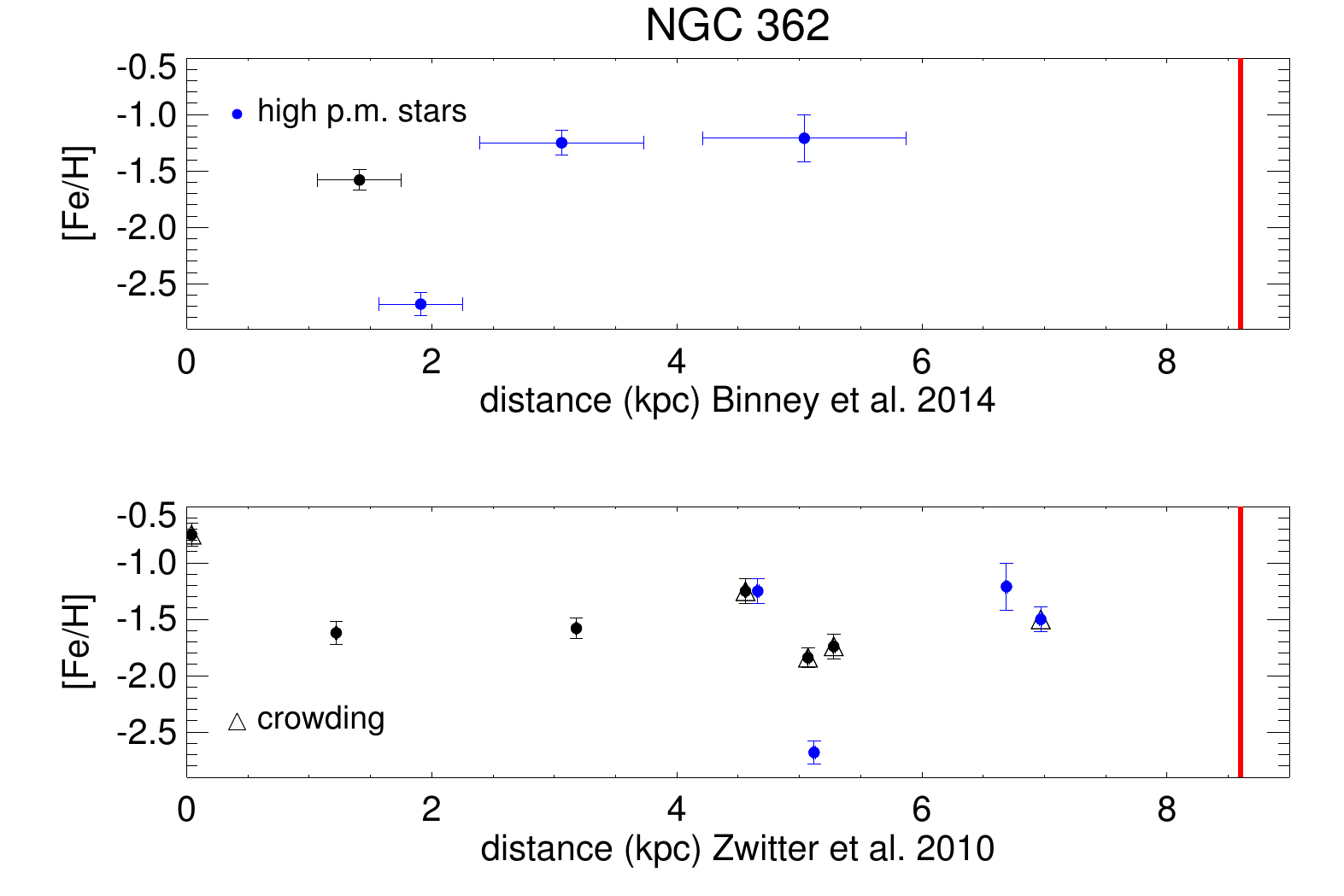}
    \caption{Distance vs. [Fe/H] for the members of NGC 362. In the top panel are plotted the distances derived in \citet{2013MNRAS.tmp.2584B}; however, there are only 4 stars with this information. In the bottom panel the distances from \citet{2010A&A...522A..54Z} are shown. Blue dots are stars with a large proper motion; triangles indicate stars affected by crowding. The vertical red line indicates the distance of the cluster from \citet{1996AJ....112.1487H} (2010 edition).}
  \label{fig:meta_dist_NGC362}
\end{figure*} 

We found that all the candidate stars associated with NGC 362 from the RAVE survey following the selection methodology described in this work have a distance significantly smaller than the nominal distance for the cluster (see Fig.~\ref{fig:meta_dist_NGC362}). However, there are only 4 stars with distance estimates from \citet{2013MNRAS.tmp.2584B}, while \citet{2010A&A...522A..54Z} estimated the distances for all the potential members.

\section{Conclusions}
 
We report the identification of potential stellar members of NGC 3201, NGC 5139 ($\omega$ Cen) and NGC 362 in the Radial Velocity Experiment-Fourth Data Release (RAVE-DR4) survey \citep{2013AJ....146..134K} using primarily the precise radial velocity derived from the RAVE spectra, for which 80$\%$ of the stars in the catalogue have $\sigma_{RV}$ $<$ 5 km s$^{-1}$. These three clusters have large systemic RVs ($\mid$RV$\mid$ $>$ 100 km s$^{-1}$), making them ideal for  relatively uncontaminated radial velocity selection from the bulk of the stellar disc. We also used proper motions (where available) and the tidal radii of the three clusters to make a reasonably robust selection of globular cluster membership. Once candidates were identified, we used them to test the precision and accuracy of the stellar parameters derived for the stars in the survey. The fact that distances and ages are known with relatively high precision for these clusters makes them ideal testbeds for these kinds of measurements in the targets observed for the survey. 

For NGC 3201 we found a star-to-star metallicity variation from -1.25 to -1.83 dex with an internal metallicity scatter of 0.24 dex. This is larger than the typical abundance errors reported in RAVE-DR4, suggesting an intrinsic metallicity scatter in the cluster. A significant spread in metallicity has also been seen in work  based on high-resolution spectroscopic observations \citep{1998AJ....116..765G,2013ApJ...764L...7S}; however, as discussed above, other authors have not found evidence for a large [Fe/H] spread (e.g. \citealt{2003PASP..115..819C,2009A&A...508..695C}), and in fact an explanation for this discrepancy may have recently been uncovered \citep{2015ApJ...801...69M}. Nevertheless, the absolute metallicity values from RAVE are in good agreement with the values reported in these studies. Using the RAVE chemical abundance catalogue \citep{2011AJ....142..193B} we found a mean [$\alpha$/Fe] $\sim$ 0.22, also consistent with high-resolution spectroscopic results. Overall there is {\em reasonable} agreement (to within the errors) between the RAVE-derived temperatures and surface gravities and the BASTI isochrones for the metallicities and ages reported in the literature for this cluster. However, if we accept the abundances from the literature, the candidate members tend to lie systematically above the relevant isochrones for ages that are generally found for globular clusters (10 - 13 Gyr), suggesting that the RAVE analysis may be underestimating surface gravities by less than 0.5 dex for these metal-poor giants. Distances to clusters can be determined very precisely; a typical uncertainty in distance for a globular cluster is $\sim$ 6$\%$ \citep{2013ApJ...775..134V}. \citet{2013MNRAS.tmp.2584B} distances found distances for the stars in our sample ranging from 2.7 to 3.7 kpc. If these objects are part of the main cluster these results would suggest that the distances for the giants we identify as members of NGC 3201 are underestimated by $\sim$ 40$\%$. \citet{2010A&A...522A..54Z} distances for stars in our sample range from 1.3 to 6.5 kpc; these distances have a significant scatter with respect to the nominal value for the cluster (D = 4.9 kpc). \citet{2013MNRAS.tmp.2584B} found an age of 10 Gyr for all the candidate members, which is only slightly different to the age of 11.4 Gyr reported by \citet{2013MNRAS.433.2006M}. 

For NGC 5139 ($\omega$ Cen) we reported a large spread in [Fe/H] ($\sim$ 2.0 dex), ranging from near solar values to [Fe/H] $\sim$ -2.2 dex for the selected candidate members. This result is in good agreement with work done by \citet{2005ApJ...634..332S}, \citet{2007ApJ...663..296V} and \citet{2010ApJ...722.1373J} using high-resolution spectroscopy. We also confirmed several star formation episodes in $\omega$ Cen from the RAVE metallicity distribution function. In addition, we found a large spread in [$\alpha$/Fe] for a given [Fe/H] using the $\alpha$-elements calculated in \citet{2011AJ....142..193B}, with [$\alpha$/Fe] ranging from -0.23 to +0.32, and a mean error of $\sim$ 0.2 dex. Large spreads in [$\alpha$/Fe] have also been found by other studies \citep{2002ApJ...568L.101P, 2007ApJ...663..296V}. As with NGC 3201, we found reasonable agreement between the T$_{\rm eff}$ and log g values derived from the RAVE spectra and the selected BASTI isochrones. The youngest stars in the cluster are believed to be around 8 Gyr \citep{2007ApJ...663..296V}. If this age limit is correct, the isochrones would suggest that the spectroscopic gravities for the most metal-poor stars are underestimated, as was also the case for NGC 3201. \citet{2013MNRAS.tmp.2584B} found most of the stars in our sample lie at distances between 2 and 4 kpc with a peak at $\sim$ 3 kpc, but with members spanning from 1 to 7 kpc. \citet{2010A&A...522A..54Z} also found stars ranging from 1 and 7 kpc; using these distances we find a group of stars with a mean value around 5 kpc and [Fe/H] between -1.0 and -2.2 dex, in good agreement with the distances for $\omega$ Cen found in the literature. 

An age spread from 8 to 13 Gyr between different stars in the cluster has been reported in the literature \citep{2007ApJ...663..296V}. However, as discussed above, other studies (e.g. \citealt{2005ApJ...634..332S, 2006ApJ...647.1075S}) have found a much smaller or even negligible age dispersion for this cluster. Unfortunately, RAVE data can only weakly constrain the ages of stars \citep{2013MNRAS.tmp.2584B}. 


For NGC 362 we found that most of the stars identified in this cluster have [Fe/H] $\sim$ -1.7 dex with a range from -0.7 to -1.8 dex. However, high resolution spectroscopic studies have measured the metallicity for this cluster at [Fe/H] $\sim$ -1.2, and have not found any spread in [Fe/H] \citep{2013A&A...557A.138C}. It is very unlikely that NGC 362 stars have such a wide range in [Fe/H], as colour-magnitude diagrams obtained for this system show very little spread (e.g. \citealt{2001AJ....122.2569B}). There is no [$\alpha$/Fe] information for these stars in the RAVE chemical abundance catalogue. The large discrepancies in both metallicity and metallicity spread between our sample of candidates and values in the literature suggest that our sample may be significantly contaminated with field stars and / or some of the RAVE-derived abundances may be erroneous. There is a general good agreement, within the errors, between potential members and the isochrones used for this cluster; however we also found that, again, using the metallicity and age reported in the literature for this cluster the isochrones suggest that the surface gravities are underestimated. Many candidates appear clearly outside the isochrones in the T$_{\rm eff}$ - log g plane, supporting the idea that these are halo field stars or SMC stars with RVs similar to NGC 362. As with NGC 3201 and $\omega$ Cen, we also found that the distances derived in the two studies based on the RAVE survey \citep{2010A&A...522A..54Z,2013MNRAS.tmp.2584B} are systematically lower than the distance reported for this cluster in the literature.

From this test of the stellar parameters, abundances and estimated distances of RAVE stars identified as potential globular cluster members, we draw two general conclusions. The first is that the derived stellar parameters and abundances are in good agreement with independent measurements based on high resolution spectroscopic studies from the literature, with the exception of surface gravities, which appear to be systematically underestimated relative to what would be predicted by BASTI isochrones for the nominal metallicities and ages of these clusters; in $\omega$ Cen, the cluster with the largest metallicity spread, this discrepancy appears to be strongest for the most metal-poor stars. 


The second general conclusion is that, assuming the candidate members are in fact associated with each cluster, and assuming that they are in the main body of each cluster (and hence at the same distance), the distances derived for these stars via two different methods are systematically low \citep[e.g.,][]{2013MNRAS.tmp.2584B}  or are distributed over a very wide range \citep[e.g.,][]{2010A&A...522A..54Z}. While the identification of some of the potential members of NGC 362 is admittedly less secure, the preponderance of evidence indicates that we have found numerous genuine members of both NGC 3201 and $\omega$ Cen in RAVE data, which, unless they are significantly extended along the line of sight, should be at approximately the same distance as their respective cluster. In both of these latter cases, the distances estimated by \citet{2010A&A...522A..54Z} appear to show a correlation with metallicity, in that the most metal-poor stars are assigned the greatest distances. Consistent with our results, an independent analysis of RAVE DR4 data suggests that the distances to metal-poor red giants from \citet{2013MNRAS.tmp.2584B} may be systematically underestimated (T. Piffl, priv. comm.). Note that \citet{2013MNRAS.tmp.2584B} assumed stellar density profiles for the thick disk and halo while \citet{2010A&A...522A..54Z} used a flat prior for stellar density vs. distance; in principle one might expect \citet{2013MNRAS.tmp.2584B} to prefer small distances, because the density of the region where a hypothetical nearby star is located would be higher and hence the overall solution for it would have a higher probability. In the case of \citet{2010A&A...522A..54Z}, no constraint on stellar density and distances could explain the larger distances scatter observed for the potential members. 

That these two general conclusions seem to be incompatible is somewhat puzzling. As noted above, one would expect systematic underestimation of surface gravities to lead to systematic overestimation of distances -- as, at a given temperature and apparent magnitude, assuming a higher luminosity places a star at a greater distance -- yet the opposite appears to be the case with the highest probability RAVE globular cluster member stars. Moreover, the possible underestimation of RAVE distances to metal-poor red giants in \citet{2013MNRAS.tmp.2584B} noted by Piffl (priv. comm.) would appear to be directly opposite to the apparent trend with metallicity seen in the distances from \citet{2010A&A...522A..54Z}. 

This conundrum notwithstanding, the work presented here demonstrates the remarkable utility of survey stars which are members of globular clusters for testing the information output by stellar survey pipelines. As astronomy reaps the rewards of an era of both current and future massive spectroscopic surveys, the validation of the stellar parameters derived from the resulting spectra is of fundamental importance. 

\section*{Acknowledgments}

Borja Anguiano and Daniel B. Zucker gratefully acknowledge the financial support of the Australian Research Council through Super Science Fellowship FS110200035 and Future Fellowship FT110100743, respectively. Borja Anguiano also thanks to Colin Navin for lively discussions of the manuscript. 

Funding for RAVE has been provided by: the Australian Astronomical Observatory; the Leibniz-Institut f\"{u}r Astrophysik Potsdam (AIP); the Australian National University; the Australian Research Council; the French National Research Agency; the German Research Foundation (SPP 1177 and SFB 881); the European Research Council (ERC-StG 240271 Galactica); the Istituto Nazionale di Astrofisica at Padova; The Johns Hopkins University; the National Science Foundation of the USA (AST-0908326); the W. M. Keck foundation; the Macquarie University; the Netherlands Research School for Astronomy; the Natural Sciences and Engineering Research Council of Canada; the Slovenian Research Agency; the Swiss National Science Foundation; the Science \& Technology Facilities Council of the UK; Opticon; Strasbourg Observatory; and the Universities of Groningen, Heidelberg and Sydney. The RAVE web site is at http://www.rave-survey.org

\appendix

\bsp

\label{lastpage}

\end{document}